\documentclass[aip,pof,reprint,onecolumn]{revtex4-1}

\usepackage{amsmath}    
\usepackage{graphicx}   
\usepackage{verbatim}   
\usepackage{color}      
\usepackage{subfigure}  
\usepackage{hyperref}   
\usepackage{tensor}

\newcommand{\Ei}[1]{\mbox{Ei}\left( {#1} \right)}
\newcommand{\er}{{\bf  {e_r}}}
\newcommand{\et}{{\bf { e_\theta}}}
\newcommand{\ez}{{\bf { e_z}}}
\newcommand{\la}{\left<}
\newcommand{\ra}{\right>}

\begin{document}

\title{Scale-dependent cyclone-anticyclone asymmetry in a forced rotating turbulence experiment}

\author{B. Gallet}
\author{A. Campagne}
\author{P.-P. Cortet}
\author{F. Moisy}
\affiliation{Laboratoire FAST, CNRS, Universit\'e Paris-Sud, B\^atiment 502, 91405 Orsay, France}

\date{\today}

\begin{abstract}
We characterize the statistical and geometrical properties of the cyclone-anticyclone asymmetry in a statistically-steady forced rotating turbulence experiment. Turbulence is generated by a set of vertical flaps which continuously inject velocity fluctuations towards the center of a tank mounted on a rotating platform. We first characterize the cyclone-anticyclone asymmetry from conventional single-point vorticity statistics. We propose a phenomenological model to explain the emergence of the asymmetry in the experiment, from which we predict scaling laws for the root-mean-square velocity in good agreement with the experimental data. We further quantify the cyclone-anticyclone asymmetry using a set of third-order two-point velocity correlations. We focus on the correlations which are nonzero only if the cyclone-anticyclone symmetry is broken. They offer two advantages over single-point vorticity statistics: first, they are defined from velocity measurements only, so an accurate resolution of the Kolmogorov scale is not required; second, they provide information on the scale-dependence of the cyclone-anticyclone asymmetry. We compute these correlation functions analytically for a random distribution of independent identical vortices. These model correlations describe well the experimental ones, indicating that the cyclone-anticyclone asymmetry is dominated by the large-scale long-lived cyclones.
\end{abstract}

\maketitle

\section{Introduction\label{sec1}}

Turbulence subject to solid-body rotation is ubiquitous in oceanic, atmospheric and astrophysical flows. Rotation  complexifies the classical phenomenology of three-dimensional turbulence by imposing a preferred direction and inducing strong anisotropy in the flow.\cite{CambonBook,DavidsonBook} This anisotropy results in two important features of rotating turbulent flows: the first one is a partial two-dimensionalization of the velocity field,\cite{Cambon1989,Staplehurst2008,Davidson2010,Lamriben2011,Sen2012,Cortet2013} and the second one is cyclone-anticyclone asymmetry.\cite{Hopfinger1982,Bartello1994,Godeferd1999,Morize2005,Bourouiba2007,Sreeni2008,Bokhoven2008,Moisy2011}

For asymptotically large rotation rates, Taylor-Proudman theorem indicates that the velocity field becomes independent of the coordinate along the rotation axis (hereafter referred to as vertical axis by convention). For intermediate rotation rates, the velocity field remains three-dimensional but has a large amount of kinetic energy in two-dimensional horizontal modes.
The quantification of two-dimensional versus three-dimensional velocity fluctuations in rotating turbulent flows is a key issue to estimate the energy dissipation rate and the mixing in such flows. Indeed, a three-dimensional turbulent flow dissipates energy at a rate per unit mass $\epsilon_{3D}$ that is independent of viscosity in the low viscosity limit: $\epsilon_{3D} \sim U^3/L$, where $U$ and $L$ are the typical velocity and length scales in the flow\cite{Frisch}. By contrast, two-dimensional turbulent flows do not present such a dissipative anomaly. They typically dissipate energy at a rate per unit mass $\epsilon_{2D} \sim \nu U^2/L^2$ at low viscosity \cite{Alexakis, Gallet}: $\epsilon_{2D}$ is smaller than $\epsilon_{3D}$ by a factor which is the inverse of a Reynolds number. One can thus expect the two-dimensionalization induced by global rotation to strongly impact the energy budget. 

Cyclone-anticyclone asymmetry is a generic feature of rotating flows, which originates from the modification of stretching and tilting of the vorticity by the Coriolis force, with an enhanced stretching of cyclonic vorticity and a selective destabilization of anticyclonic vorticity. Starting from homogeneous, isotropic and parity-invariant turbulence, adding rotation along the vertical axis breaks the invariance to any reflection $\mathcal{S}_\parallel$ with respect to a vertical plane: there is a preferred direction of rotation in the flow. This results in an asymmetric vertical vorticity distribution, with an enhanced probability for large cyclonic vorticity values (vertical vorticity of the same sign as the background rotation). A visual consequence of this asymmetry is the predominance of cyclones over anticyclones in rotating turbulent flows, hence the denomination cyclone-anticyclone asymmetry. Here we call `cyclone-anticyclone asymmetric' a turbulent flow which has a nonzero value for any statistical quantity that would be zero in the presence of $\mathcal{S}_\parallel$ symmetry.

Cyclone-anticyclone asymmetry in homogeneous rotating turbulence has received considerable interest in recent years. It has been first characterized experimentally from visualizations of forced rotating turbulence,\cite{Hopfinger1982} and further analyzed quantitatively in a series of numerical and experimental studies, including forced\cite{Godeferd1999} and freely decaying turbulence.\cite{Bartello1994,Morize2005,Bourouiba2007,Sreeni2008,Bokhoven2008}
However, all these studies focused on single-point vorticity statistics, such as the vorticity skewness, without addressing the important issue of the scale dependence of the asymmetry. 
Here the vorticity skewness is defined as $S_\omega=\left< \omega^3 \right>/ \left< \omega^2 \right>^{3/2}$, where the angular brackets denote ensemble average and $\omega$ is the vorticity component along the rotation axis. This quantity vanishes in a turbulent flow with $\mathcal{S}_\parallel$ symmetry, and is positive in homogeneous rotating turbulence. Several fundamental questions thus remain to be answered: knowing the statistical properties of a non-rotating turbulent flow, can one predict the amount of cyclone-anticyclone asymmetry that arises as a consequence of adding global rotation to the system? What scales of the turbulent flow contain this asymmetry? Does the asymmetry first appear at small scales, or is it dominated by a few large-scale cyclones?

A first attempt to characterize the scale-dependence of cyclone-anticyclone asymmetry has been proposed by Moisy {\it et al.}\cite{Moisy2011}. They computed the skewness of the coarse-grained vertical vorticity field, and they observed that the filter size for which $S_\omega$ is maximum is comparable to the size of the large-scale cyclones in the flow. In the present paper we extend this approach using third-order antisymmetric velocity correlations, which provide a systematic scale-by-scale quantification of cyclone-anticyclone asymmetry.  In the asymptotic regime of very large Reynolds number with moderate to small Rossby number, these third-order correlations are good candidates to characterize a self-similar distribution of cyclone-anticyclone asymmetry between the different scales of rotating turbulent flows. Furthermore, these correlations are simpler to measure than vorticity statistics. Indeed, they can be computed from velocity measurements on a rather coarse spatial grid, from two-point velocity measurements, or even from one-point velocity measurements provided Taylor's hypothesis can be applied: they are thus well-suited to quantify cyclone-anticyclone asymmetry in oceanographic and atmospheric data. For instance, related quantities (antisymmetric third-order velocity structure functions) were computed by Lindborg and Cho\cite{LindborgCho} in an analysis of atmospheric data, but they did not relate them to the physics of cyclone-anticyclone asymmetry.

Analytical results on cyclone-anticyclone asymmetry are mostly of two kinds: first, linear stability analyses indicate that anticyclones are more subject to the centrifugal instability than cyclones.  Indeed, for an axisymmetric vortex in an rotating frame, Rayleigh's criterion for stability of the vortex with respect to axisymmetric perturbations can be expressed in terms of the monotony of the total angular momentum profile (see for instance Ref.~\onlinecite{Kloosterziel}). The resulting criterion indicates that cyclones are stable, whereas anticyclones with moderate local Rossby number are centrifugally unstable. As a consequence of this instability the anticyclones break into  three-dimensional motions that are effectively damped by viscosity. Interestingly, for rapid global rotation, that is for low absolute value of the local Rossby number, the anticyclones become centrifugally stable.

The second kind of analysis on cyclone-anticyclone asymmetry addresses fully turbulent flows. Gence and Frick\cite{Gence} considered homogeneous isotropic turbulence that is suddenly subjected to the Coriolis force at $t=0$, and studied the subsequent evolution of the velocity field for early times. This approach is detailed and extended using rapid-distortion theory (RDT) in Ref.~\onlinecite{Bokhoven2008} (see Ref.~\onlinecite{Cambon} for details on RDT). The thought experiment of Gence and Frick can be realized numerically by solving the non-rotating Navier-Stokes equation for some time, before adding the Coriolis force to the equation at some arbitrary time. However, in an experiment, a brutal increase in the rotation rate from zero to some finite value would induce a strong Poincar\'e force in the rotating frame that is absent from the analysis of Gence and Frick. Nevertheless, these authors show that, for early times, the third moment of the distribution of vertical vorticity $\omega$ grows linearly in time, at a rate proportional to $\Omega$, and to the enstrophy production rate due to vortex stretching. Denoting still with square brackets the ensemble average, this translates into the following approximate balance for short time
\begin{equation}
\left. \frac{\mathrm{d}}{\mathrm{d}t} \left< \omega^3 \right> \right|_{t=0^+} \sim \Omega \left. \left< \omega^2 \right>\right|_{t=0} ^{3/2} \, \label{GF}.
\end{equation}
If we consider decaying turbulence for $t>0$, the decay time of the small scales that contribute to $\omega$ is roughly the Kolmogorov timescale $\left<\omega^2 \right>^{-1/2}$. Assuming that Eq.~(\ref{GF}) remains valid during this decay time, the third-order vorticity moment is therefore given by
\begin{equation}
\left< \omega^3 \right>  \sim \Omega \left< \omega^2 \right> \, , \label{ODGo3}
\end{equation}
or in terms of the vorticity skewness, $S_\omega \sim \Omega / \langle \omega^2 \rangle^{1/2}$.
This order of magnitude can be thought of as a Taylor expansion in the weak rotation limit: the vorticity asymmetry vanishes without global rotation and is proportional to $\Omega$ for weak rotation rates. Such an order of magnitude is of limited predictive power, and a precise quantification of cyclone-anticyclone asymmetry in forced rotating turbulence thus remains to be performed.

In the following we characterize the scale-dependence of the cyclone-anticyclone asymmetry from particle image velocimetry (PIV) measurements performed in a forced rotating turbulence experiment. The setup and PIV system are mounted on a rotating platform, so we can add global rotation to the system and study how it affects the turbulence once a statistically steady-state is reached. This experimental setup is presented in section \ref{sec2}. In section \ref{sec3}, we characterize the cyclone-anticyclone asymmetry using conventional single-point vorticity statistics. We propose a phenomenology for the emergence of this asymmetry. This phenomenology leads to scaling laws for the root-mean-square velocity that we confront to the experimental data. In section \ref{sec4} we introduce a set of third-order two-point velocity correlation functions to quantify the cyclone-anticyclone asymmetry, from which the scales dominating the asymmetry can be determined. To get some insight on what these correlations represent, we compute them for a random distribution of two-dimensional vortices in section \ref{sec5}. The good agreement with the experimental correlation functions indicates that the asymmetry mostly originates from a few large-scale long-lived vortices in the PIV domain.


\section{Experimental setup\label{sec2}}

\begin{figure}
 \subfigure[]{\label{expsetupa}
\includegraphics[height=55 mm]{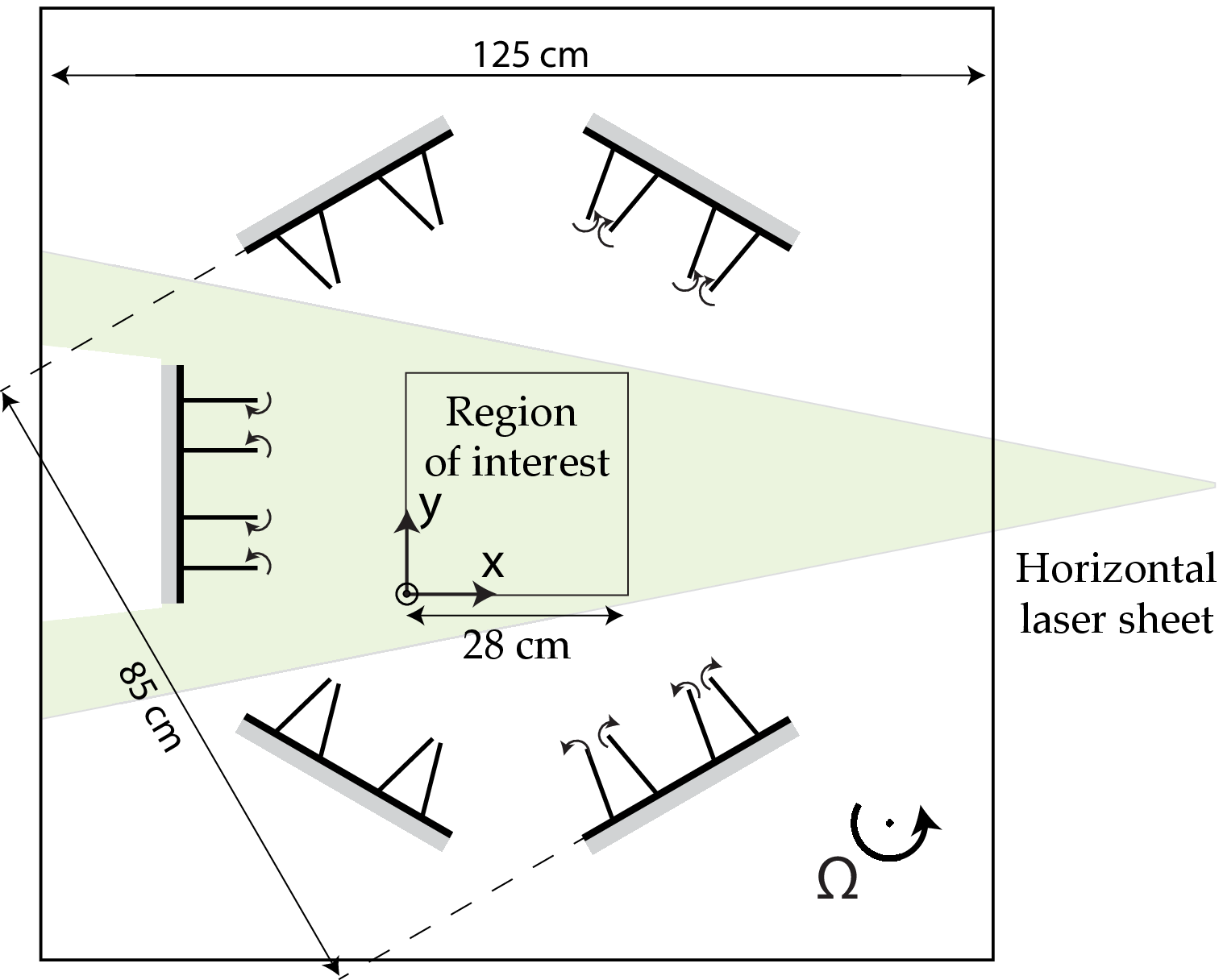}
}\hspace{15 mm}
 \subfigure[]{\label{expsetupb}
\includegraphics[height=55 mm]{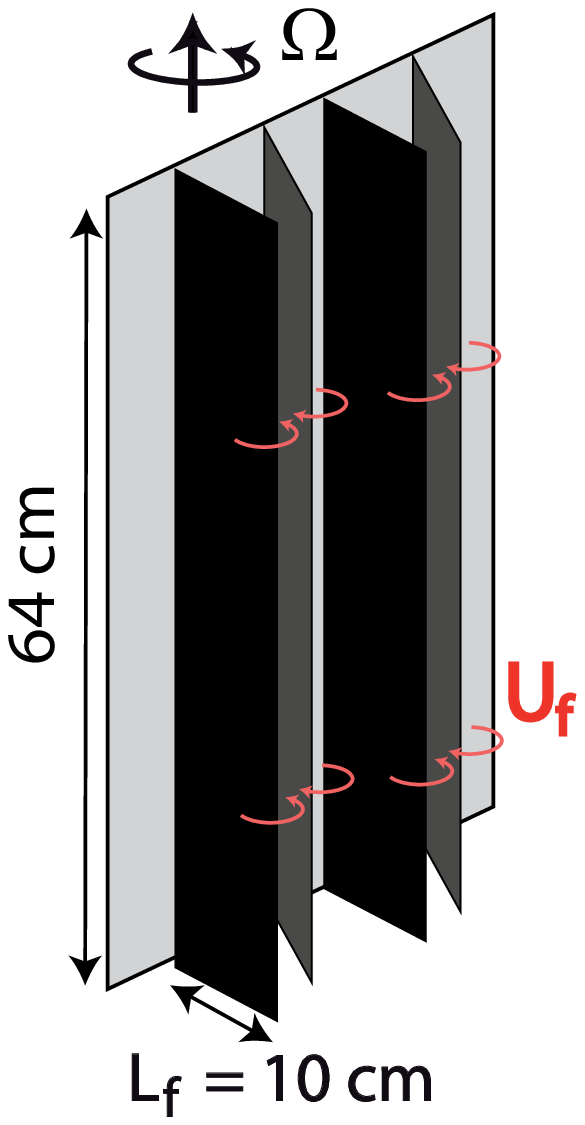}
}
\caption{(a) Top-view of the experimental setup. An arena of 10 pairs of flaps forces a turbulent flow in the central region of a square-based tank. We perform PIV measurements in a horizontal domain in the central region of the tank. The tank and PIV systems are mounted on a rotating platform that spins around a vertical axis with constant angular velocity $\Omega$. (b) Schematics of a vertical block  with two pairs of flaps. During one period of motion a pair of flaps closes rapidly and generates three-dimensional velocity perturbations that propagate toward the central region of the tank, before opening again. The tips of the flaps have linear velocity $U_f$ during this closing phase.} 
\label{expsetup}
\end{figure}

The experimental setup is presented in Fig.~\ref{expsetup}: a water-filled tank with a square base of $1.25 \times 1.25$~m$^2$ area and a depth of $0.60$~m is mounted on a precision two-meter-diameter rotating platform. The angular velocity $\Omega$ of the platform ranges from $0$ to $20$~revolutions per minute (rpm), with less than $10^{-3}$ relative fluctuations. We produce statistically steady turbulence in the central region of the tank using 10 pairs of vertical flaps organized in an approximately circular arena of diameter $85 \pm 5$ cm (Fig.~\ref{expsetupa}). This forcing procedure was originally developed at LadHyX (Ecole Polytechnique) to generate turbulence in stratified fluids.\cite{Billant,Augierthese,Augier} A detailed view of the flaps is presented in Fig.~\ref{expsetupb}: the flaps have a horizontal length $L_f=100$~mm.  Each one of them is attached with hinges to a vertical block so it can rotate around a vertical axis. We impose to each pair of flaps the following periodic motion of period $T$:  they open during a time $0.2\, T$ until they are parallel one to the other. They remain parallel and steady during $0.2\,T$, before closing during $0.2\,T$. At the end of the closing motion, the tips of the two flaps are approximately $15$ mm apart. During the closing motion, each flap describes an angular displacement of approximately $9^\circ$ around its vertical axis.
The two flaps remain steady during the last $0.4\,T$ of the period, before opening again at the beginning of the next period. Each pair of flaps follows the same periodic sequence but the five blocks seen in Fig.~(\ref{expsetupa}) have random phase differences. 

We denote as $U_f$ the  velocity of the tip of a flap during the closing motion (prescribing $U_f$ is equivalent to specifying the period $T$ of the motion of the flaps, since the former is inversely proportional to the latter). In the following, the flap velocity $U_f$ ranges between $4.2$ and $9.2$ mm~s$^{-1}$. The closing of a pair of flaps expells the fluid between the flaps towards the central region of the tank. If the motion of the flaps is slow enough, this produces a vortex dipole that self-propagates towards the center of the tank\cite{Augier,Augierthese,Billant}. However, in the present experiment the Reynolds number based on the flap velocity is large: without global rotation, the flow close to the flaps rapidly becomes turbulent and three-dimensional, with a mean horizontal velocity directed towards the center of the arena and strong three-dimensional fluctuations. The periodic closing of the flaps thus produces turbulent ``bursts" that propagate towards the central region of the tank. By contrast, in the regime of asymptotically large rotation rate, these vertically-invariant flaps allow for a Taylor-Proudman regime, where the flow is independent of the vertical coordinate (away from the top and bottom boundaries). Note that the same forcing technique applied to a stably stratified fluid leads to a very different situation: the vortex dipoles are subject to a strong zig-zag instability and the flow organizes into horizontal layers.\cite{Billant,Augierthese}

\begin{table}[bt]
\begin{center}
\begin{tabular*}{0.75\textwidth}{@{\extracolsep{\fill}} c  c  c  c  c  c  c}
$\Omega$ [rpm] & $Ro$ & $u_\text{rms}$ [mm/s] & $L_\text{int}$ [mm] & $Re^{(l)}$ & $Ro^{(l)}$ & $\tau$\\
\hline
0 & $\infty$ & 7.4 & 37 & 276 & $\infty$ & 0.66 \\
4 &  0.11   &  8.5   & 89     &  760    &  0.11     &  0.88 \\
8 &  0.055   &  11.8   & 88     &  1040    &  0.080     & 0.97 \\
12 &  0.037   &  13.1   & 78     &  1020    &  0.067     & 0.97 \\
16 &  0.028   &  14.6   & 76     &  1110    &  0.057     &  0.98 \\
20 &  0.022   &  15.2   & 73     &  1110    &  0.050     &  0.98
\end{tabular*}
\caption{Global quantities in the turbulent flow for various rotation rates, given for the maximum flap velocity, $U_f=9.2$ mm~s$^{-1}$, which corresponds to a control Reynolds number $Re=920$.
\label{table1}}
\end{center}
\end{table}

Each run is characterized by two control parameters: the angular velocity $\Omega$ of the platform and the flap velocity $U_f$. The system is therefore governed by two independent dimensionless quantities, the Reynolds and Rossby numbers, defined as
$$
Re= U_f L_f /\nu, \qquad Ro= U_f / (2 \Omega L_f).
$$
For the maximum value of the flap velocity, $U_f=9.2$ mm~s$^{-1}$, one has $Re = 920$, while $Ro$ ranges between $\infty$ in the absence of rotation to $0.022$ for the maximum rotation rate $\Omega=20$ rpm (see Table \ref{table1}).

\begin{figure}
 \subfigure[]{\label{snapshotsa}
\includegraphics[width=65 mm]{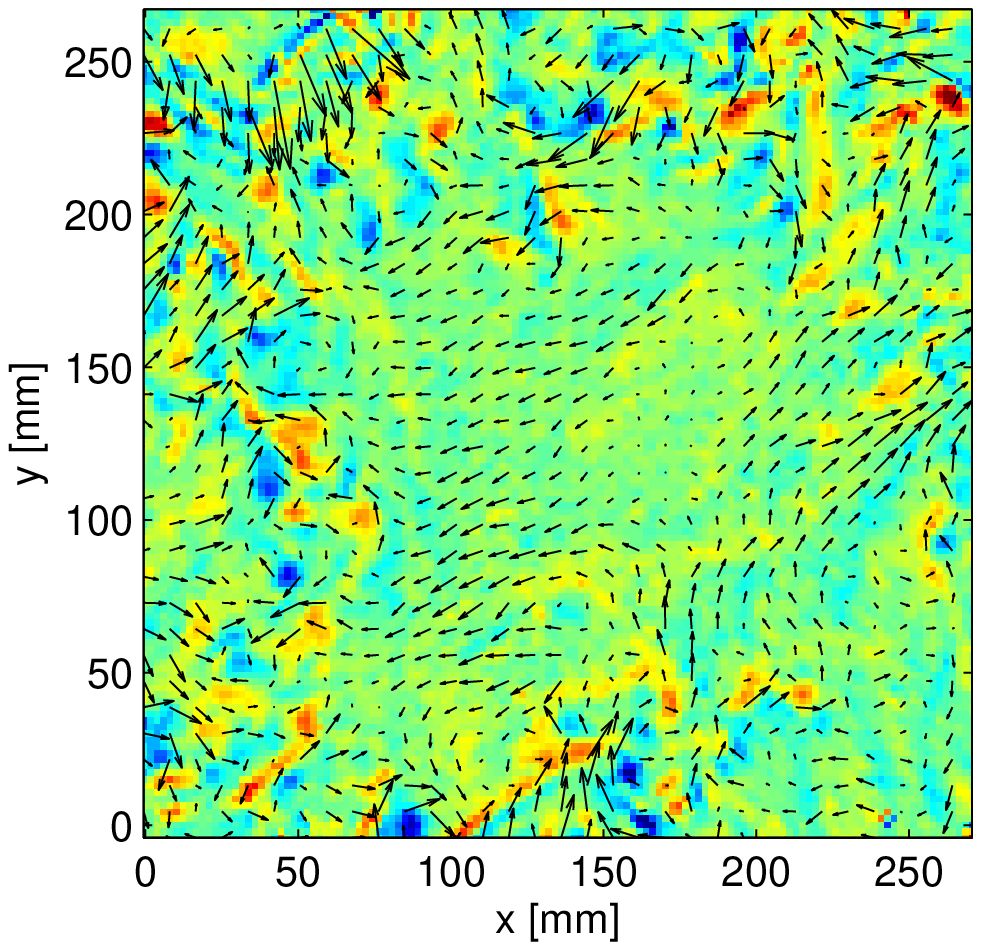}
}
 \subfigure[]{\label{snapshotsb}
\includegraphics[width=65 mm]{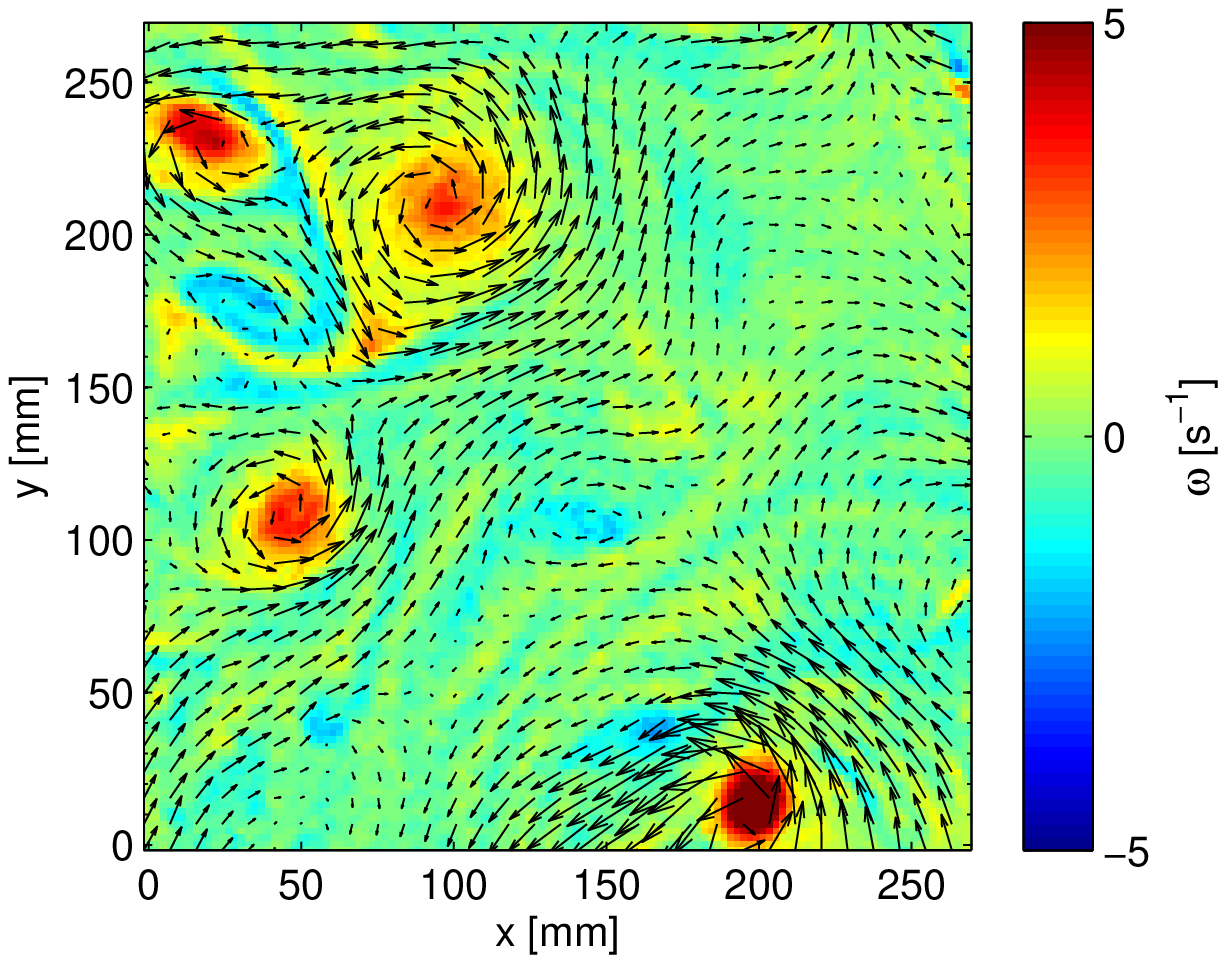}
}
\caption{Snapshots of the PIV measurements in the statistically-steady state for $U_f=9.2$ mm/s. The arrows are the fluctuating part of the velocity field and color is the corresponding PIV vorticity. (a) Non-rotating case. One observes small-scale turbulent fluctuations travelling from the flaps to the central region of the tank. (b) Rapidly rotating case ($\Omega=16$ rpm). One can see mostly large-scale cyclones: rotation breaks the $\mathcal{S}_\parallel$ reflection symmetry, which leads to an asymmetry between cyclones and anticyclones in the turbulent rotating flow.}
\label{snapshots}
\end{figure}

After a statistically-steady state is reached, we measure horizontal velocity fields in the rotating frame using a corotating particle image velocimetry (PIV) system. The measurement domain is a $28 \text{ cm}\times 28 \text{ cm }$ horizontal square at mid-depth inside the tank.   We acquire up to 30\,000 image pairs at 1~Hz using a double-frame $2048^2$ pixel camera (large data sets are needed to ensure good convergence of the third-order correlations). Horizontal velocity fields are computed on a $128 \times 128$ grid with resolution $\ell_{\text{piv}}=2.1$ mm.
The first step of the data analysis consists in extracting the fluctuating part from the PIV velocity field $\tilde{\textbf{u}}(x,y,t)$: for given $U_f$ and $\Omega$ we first compute the time-averaged velocity field $\left< \tilde{\textbf{u}} \right>_t (x,y)$ by averaging all the PIV velocity fields (we assume ergodicity in the statistically steady state, so that ensemble average $\left< \dots \right>$ is equivalent here to time average). 
We subtract this average to all instantaneous velocity fields, to obtain the fluctuating, or ``turbulent" part of the velocity field $\textbf{u}(x,y,t)= \tilde{\textbf{u}}(x,y,t)-\left< \tilde{\textbf{u}} \right>_t(x,y)$. 
As an example, in Fig.~\ref{snapshots} we show snapshots of the fluctuating part of the PIV fields measured in the statistically-steady state for non-rotating and rotating experiments, both for a flap velocity $U_f=9.2$ mm~s$^{-1}$.

The turbulent rms velocity is defined as $u_\text{rms}=\left<  \textbf{u}^2 \right>^{1/2}_{\textbf{x},t}$, where the average is performed over space and time.  When the platform rotates the total velocity field is dominated by its fluctuating part, with negligible time-averaged flow. This can be evaluated by the turbulence rate $\tau=\left< \textbf{u}^2 \right>_{\textbf{x},t}/\left< \tilde{\textbf{u}}^2 \right>_{\textbf{x},t}$, which is 0.66 without rotation, but is at least 0.95 for $\Omega \geq 8$ rpm (see Table \ref{table1}).

In the presence of rotation, turbulence in the PIV domain can be considered approximately statistically homogeneous and axisymmetric. More precisely, the turbulent kinetic energy has a minimum at the center of the arena: the turbulent bursts injected by the closing flaps decay as they propagate towards the center of the arena. Without rotation, the turbulence is strongly three-dimensional and efficiently dissipates kinetic energy during this propagation. As a result the turbulent fluctuations are stronger at the periphery of snapshot \ref{snapshotsa} than at its center. By contrast, when the platform rotates the turbulence becomes quasi-two-dimensional and dissipates energy much more slowly, which results in much better homogeneity (see snapshot \ref{snapshotsb}): the time-averaged turbulent kinetic energy has 30\% rms spatial variations over the PIV domain for $\Omega = 0$, but these variations decrease down to 10\% for $\Omega \geq 8$ rpm.

The typical size of the energy-containing eddies in the PIV domain is characterized by the integral scale $L_\text{int}$, which is computed from the normalized longitudinal correlation function of the fluctuating horizontal velocity field. The normalized correlation function is integrated from $r=0$, where it is unity, up to the scale at which it falls under $0.05$ (we checked that doubling this arbitrary threshold does not change $L_\text{int}$ by more than 10\%). As soon as the rotation rate is above $2$ rpm, $L_\text{int}$ takes values in the range $70$ to $90$ mm, i.e. it is comparable to the flap size $L_f$. By contrast, $L_\text{int}$ is significantly smaller without rotation, $L_\text{int} \simeq 37$~mm. This traces back to a transition between purely 3D turbulence when rotation is almost zero, to mixed 2D and 3D turbulence for higher rotation rates. This transition is described in details in section \ref{phenom}. Turbulence in the center of the arena can be alternatively characterized by the local (or turbulent) Reynolds and Rossby numbers, 
$$
Re^{(l)}= u_\text{rms} L_\text{int} /\nu, \qquad Ro^{(l)}=u_\text{rms} / (2 \Omega L_\text{int}),
$$
based on the integral length $L_\text{int}$ and the horizontal rms velocity $u_\text{rms}$. Although these local and global Reynolds and Rossby numbers are of the same order of magnitude (see Table \ref{table1}),  the local numbers are not simply proportional to the global ones because global rotation modifies the turbulence generation mechanism in the vicinity of the flaps and the subsequent turbulence decay towards the center of the arena.

\section{Cyclone-anticyclone asymmetry\label{sec3}}

\subsection{Single-point vorticity statistics}

\begin{figure}
 \subfigure[]{
\includegraphics[width=65 mm]{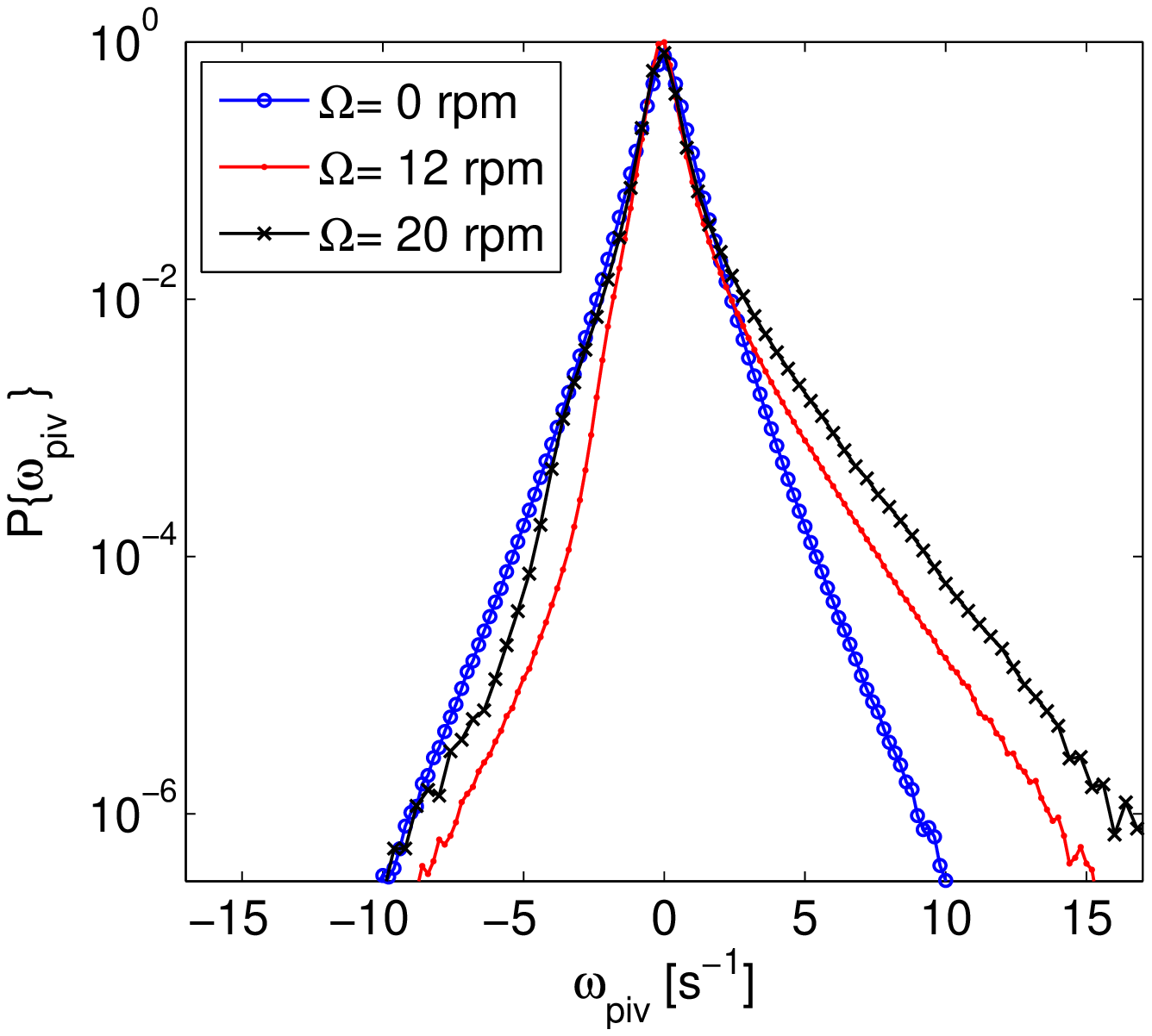}
}
 \subfigure[]{
\includegraphics[width=65 mm]{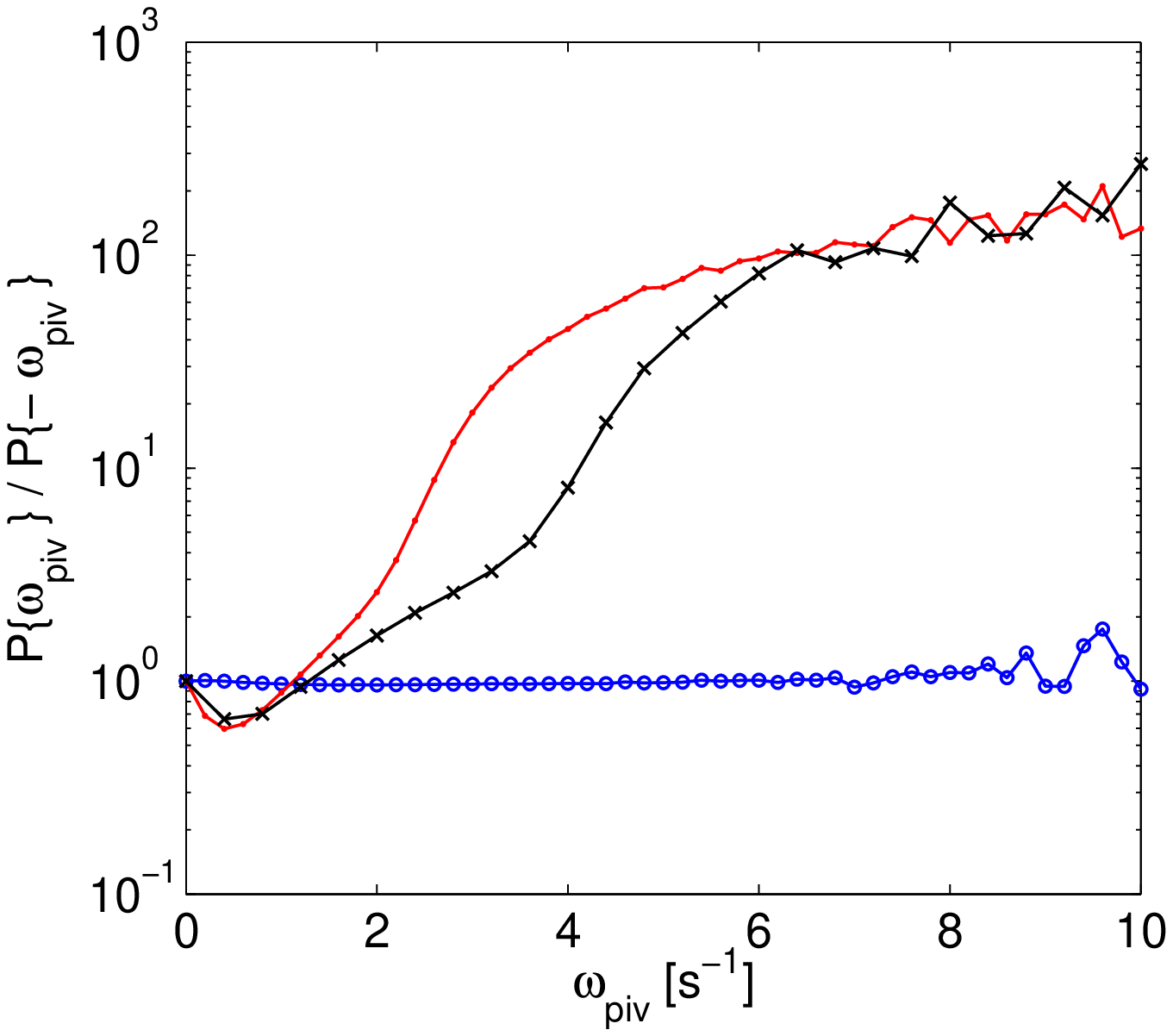}
}
 \subfigure[]{
\includegraphics[width=65 mm]{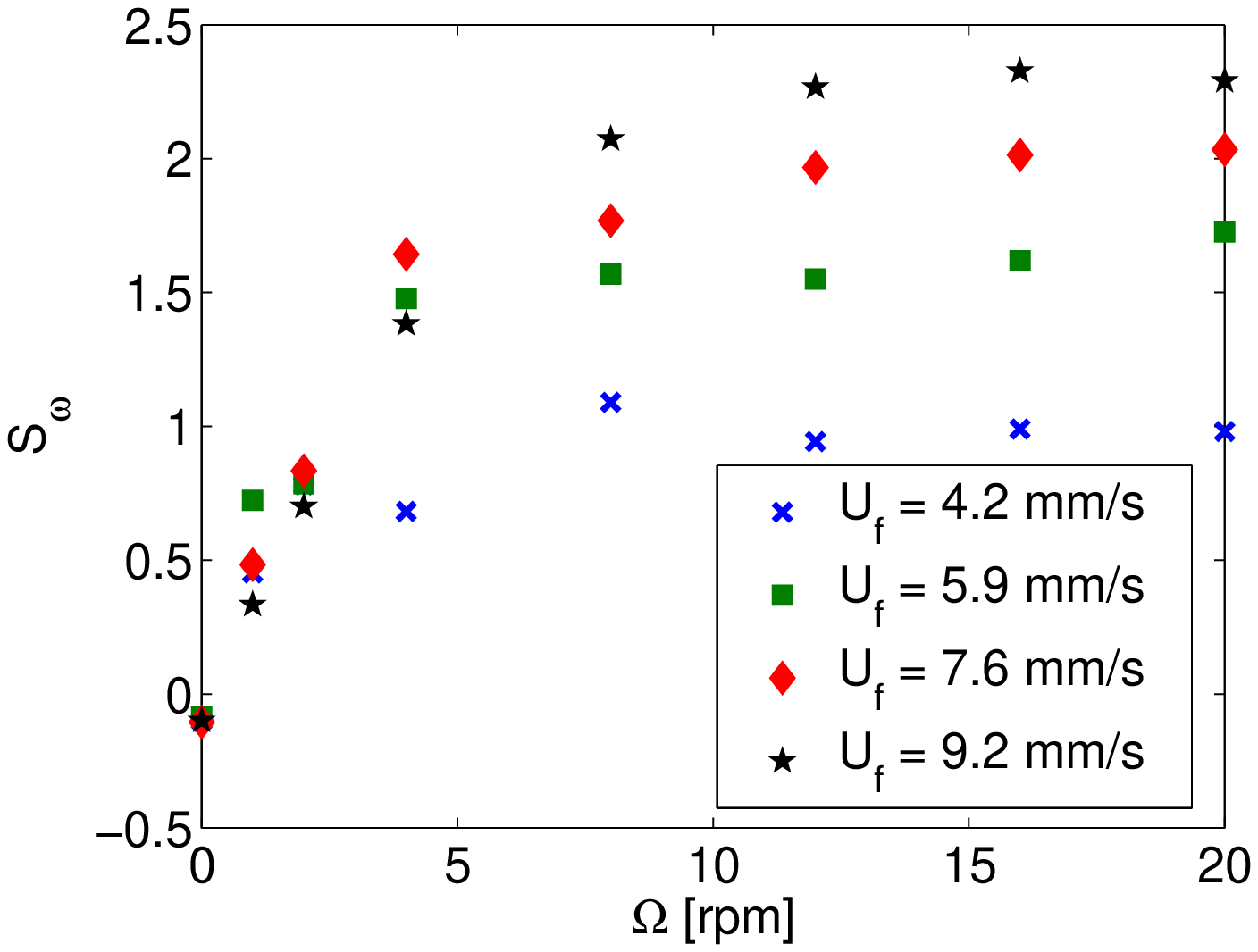}
}
\caption{(a) PDF of the vorticity obtained through finite-differences on the PIV grid. This PDF is asymmetric for nonzero global rotation. (b) Ratio of the positive over negative sides of the PDF. This ratio is close to unity without rotation, indicating symmetry of the PDF. For rapid rotation, it reaches approximately 100 for large vorticity, indicating the preference for strong cyclonic vorticity.  (c) Vorticity skewness as a function of the rotation rate, for several flap frequencies. It increases with global rotation and saturates in the limit of rapid rotation.\label{statvort}}
\end{figure}

In the snapshots \ref{snapshotsa} and \ref{snapshotsb} we show the vertical PIV-vorticity $\omega_{\text{piv}}$: 
the true vorticity is approximated using standard second-order finite differences on the PIV grid. Because the resolution  $\ell_{\text{piv}}$ of the PIV grid is larger than Kolmogorov scale $\eta$ in the experiment, the smallest scales of the turbulence are not resolved, so the computed vorticity is a coarse-grained version of the true vorticity. More precisely, in the non-rotating case the Kolmogorov scale is $\eta=(\nu^3 / \epsilon)^{1/4}$, where the energy dissipation rate per unit mass is evaluated through $\epsilon \simeq U_f^3 / L_f$. For the maximum flap velocity $U_f=9.2$ mm/s, we obtain $\eta \simeq (\nu^3 L_f/ U_f^3)^{1/4}\simeq0.6$ mm, that is $\ell_{\text{piv}} / \eta \simeq 3.5$. In spite of this limitation, this coarse-grained vorticity on a scale $\ell_{\text{piv}} $ is useful to identify rotating structures (cyclones and anticyclones) in the experiment. 

For the non-rotating case, the snapshot \ref{snapshotsa} displays small-scale vorticity fluctuations corresponding to the three-dimensional turbulent bursts emitted by closing the flaps. By contrast, in the rotating case a few large-scale coherent structures dominate the vorticity field (see \ref{snapshotsb}): this traces back to the two-dimensionalization of turbulent flows under rapid rotation, with reduced direct energy transfers towards small scales and the emergence of long-lived large-scale structures.

There is a clear predominance of cyclones over anticyclones, that traces back to the broken reflection symmetry $\mathcal{S}_\parallel$ with respect to a vertical plane. We checked that cyclones with negative vorticity dominate the flow when the platform is rotated in the opposite direction. A straightforward way to quantify this asymmetry is to compute the probability density function (PDF) of the vertical vorticity  in the statistically steady state. To compute this pdf we gather the values of the vorticity at every time and every point of the PIV fields. Such PDFs are displayed in Fig.~\ref{statvort} for several rotation rates, with constant flap velocity $U_f=9.2$ mm~s$^{-1}$. They exhibit approximately exponential tails on both sides. Without rotation the PDF is symmetric with respect to zero, whereas with rotation $\Omega>0$ the positive tail becomes dominant over the negative one. This phenomenon is highlighted in fig.~\ref{statvort}(b), where we plot the ratio of the right-hand-side ($\omega_{\text{piv}}>0$) over the left-hand-side ($\omega_{\text{piv}}<0$) of the PDF. This ratio is constant and close to unity without rotation, whereas it increases up to values of the order of $100$ for large vorticity under rapid rotation.

A classical way to quantify the asymmetry of the vorticity distribution is to introduce the vorticity skewness  $S_\omega=\left< \omega_\text{piv}^3 \right>/ \left<  \omega_\text{piv}^2 \right>^{3/2}$, which we plot in Fig.~\ref{statvort} for several rotation rates $\Omega$ and flap velocities $U_f$. For a given flap velocity, $S_\omega$ increases from zero without rotation up to values of the order of $2$ for rapid rotation. These values are comparable or larger than the ones typically found in decaying rotating turbulence experiments\cite{Morize2005,Staplehurst2008,Moisy2011} and simulations\cite{Bartello1994,Bokhoven2008}. The vorticity skewness seems to saturate for the largest $\Omega$ achieved in the experiment, the value at saturation depending on $U_f$.

\subsection{Phenomenology for the emergence of the cyclone-anticyclone asymmetry \label{phenom}}

\begin{figure}
\begin{centering}
 \subfigure[]{
\includegraphics[height=60 mm]{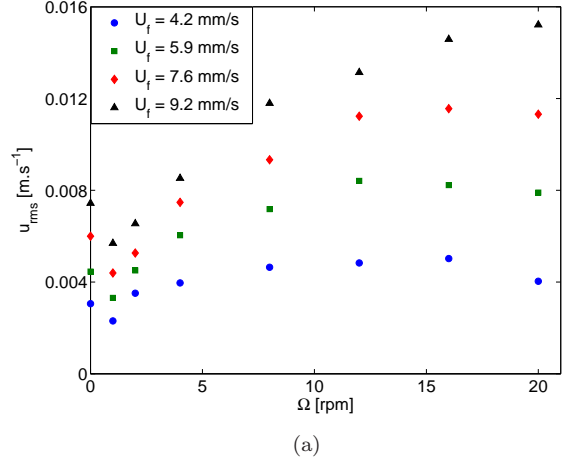}
}
 \subfigure[]{
\includegraphics[height=60 mm]{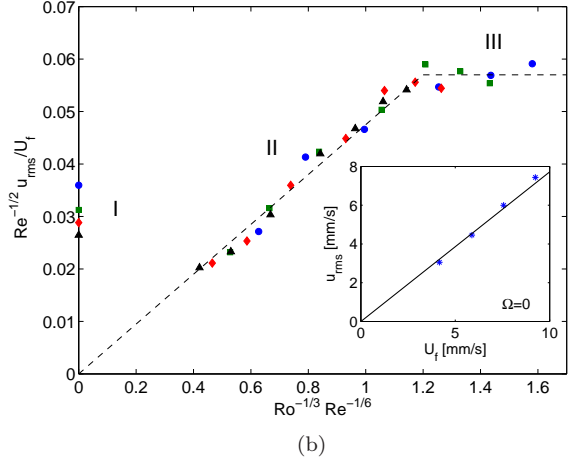} \label{figurmsb}
}
\caption{(a) Root-mean-square turbulent velocity fluctuations in the measurement domain, as a function of global rotation rate, for several flap velocities. (b) Rescaled rms velocity. The collapse of the different curves onto a single master curve confirms the different scaling regimes. The points measured in absence of rotation follow the  ``fully 3D" scaling regime (I): the inset highlights the linear relation between the rms velocity and the flap velocity when $\Omega=0$. The dashed linear law corresponds to the slow rotation scaling regime (II). The dashed plateau for faster rotation corresponds to the scaling regime of rapid global rotation (III). The legend gives the value of $U_f$ and applies to both panels.}
\label{figurms}
\end{centering}
\end{figure}

Since rotation has no effect on strictly vertically invariant flows (this is because the Coriolis force can be fully balanced by the pressure gradient), the vorticity asymmetry must emerge from an intermediate 2D-3D state. In the present configuration, the emergence of cyclone-anticyclone asymmetry can be related to the turbulence generation process: when a pair of flaps closes, it emits a turbulent burst toward the center of the arena. For moderate global rotation, this burst contains strong three-dimensional velocity fluctuations. During the propagation of this three-dimensional turbulence, rotation tends to two-dimensionalize the flow while favoring cyclonic vorticity, since cyclones are more stable than anticyclones. For rapid global rotation, this two-dimensionalization is very efficient: we have checked, using complementary PIV measurements in a vertical plane, that turbulence in the central region of the arena is indeed approximately two-dimensional for $\Omega \geq 2$ rpm.


From this scenario we can estimate the turbulent kinetic energy in the flow as a function of the parameters $Ro$ and $Re$. In the following we derive scaling laws for the root-mean-square velocity $u_{\text{rms}}$. Three regimes can be identified, going from slow to rapid global rotation:
\begin{itemize}
\item Regime I: for zero or very weak global rotation, rotation should play no role. This is a fully 3D regime, with $u_{\text{rms}}$ scaling like the flap velocity $U_f$.
\item Regime II: for moderate global rotation (large but finite $Ro$), we assume that the vorticity asymmetry is linear in $\Omega$. More precisely, the 3D turbulent bursts decay rapidly, on a time scale $L_f/U_f$, during which rotation converts a small fraction of the initial kinetic energy of the burst into large-scale two-dimensional motion with cyclone-anticyclone asymmetry. In the introduction, we stressed the fact that two-dimensional turbulence decays much more slowly than three-dimensional turbulence, with a decay rate of kinetic energy that is proportional to viscosity and thus very low in the large-Reynolds-number limit. Hence, in the central region of the tank, the rms velocity is dominated by the long-lived, nearly two-dimensional, and cyclone-anticyclone asymmetric flow, while the three-dimensional fluctuations have essentially decayed to zero. This approach is similar to that of Gence and Frick:\cite{Gence} generating a 3D turbulent burst in a rotating fluid environment may be seen as a good experimental realization of their assumption that the Coriolis force suddenly ``appears" at some initial time. We thus evaluate the vorticity asymmetry through equation (\ref{ODGo3}). In this equation, $\left<  \omega^2 \right>$ should be thought of as the enstrophy in the 3D turbulent burst, where rotation generates cyclone-anticyclone asymmetry, whereas $\left<  \omega^3 \right>$ is the final vorticity asymmetry in the 2D flow observed in the central region of the tank. Hence $\left<  \omega^2 \right>$ follows the usual 3D scaling,  $\left<  \omega^2 \right> \sim \epsilon/\nu$, where the energy dissipation per unit mass $\epsilon$ scales like $U_f^3 / L_f$.\cite{Frisch}  By contrast, $\left<  \omega^3 \right>$ should be evaluated for the two-dimensional flow, and is simply $u_{\text{rms}}^3/L_f^3$. Equation (\ref{ODGo3}) then gives
\begin{equation}
\frac{u_{\text{rms}}^3}{L_f^3} \sim \Omega \frac{U_f^3}{\nu L_f}, \qquad \mbox{hence } \quad  u_{\text{rms}} \sim U_f \left( \frac{Re}{Ro} \right)^{1/3} \, . \label{urmsregime1}
\end{equation}
In this second regime, the rms velocity should therefore scale like $\Omega^{1/3}$ for constant $U_f$.
\item Regime III: 
for rapid global rotation ($Ro \ll 1$), two-dimensionalization is very efficient, and the power injected by the flaps into the fluid is dissipated by 2D turbulence. During each period, the flaps give a typical velocity $U_f$ to the fluid in their vicinity. The fluid thus receives a kinetic energy per unit mass proportional to $U_f^2$ during each flap period $T \sim L_f / U_f$, hence a power injection per unit mass of order $U_f^3 / L_f$. Two-dimensional turbulence in the central region of the tank dissipates energy at a typical rate $\nu \,u_{\text{rms}}^2/L_f^2$. The balance between this dissipation and the injection of energy by the flaps leads to
\begin{equation}
\frac{U_f^3}{L_f} \sim \nu \frac{u_{\text{rms}}^2}{L_f^2}, \qquad \mbox{hence } \quad u_{\text{rms}} \sim U_f \sqrt{Re} \, .  \label{urmsregime2}
\end{equation}
Interestingly, in this rapid rotation regime the rotation rate does not appear anymore in the scaling law for the rms velocity, as opposed to (\ref{urmsregime1}).
\end{itemize}

Note that the transition between regimes I and II occurs for very small rotation rates. Indeed, this transition happens when the 3D velocity $U_f$ becomes smaller than the 2D rms velocity given by equation (\ref{urmsregime1}), that is as soon as $Ro < Re$. The corresponding transition rotation rate is small, $\Omega \sim \nu/L_f^2=10^{-4}$ rad/s for water. The transition between regimes II and III can be obtained by equating (\ref{urmsregime1}) and (\ref{urmsregime2}), which yields $Ro^2 Re \simeq 1$. This criterion was derived in \onlinecite{Canuto} and \onlinecite{Mininni} as the threshold for which the entire inertial range is influenced by global rotation. Indeed, if initially isotropic turbulence is suddenly subjected to global rotation, the large scales will become anisotropic, from the integral scale down to the Zeman scale, $l_Z=\sqrt{\epsilon/\Omega^3}$. If $l_Z$ is smaller than the Kolmogorov length $\eta$, then the entire inertial range is directly influenced by global rotation. Using $\epsilon=U_f^3/L_f$, the criterion $l_Z<\eta$ is equivalent to $Ro^2 Re < 1$. The present experiment also suggests an alternate explanation for this criterion: let us consider a pair of closing flaps in an initially steady fluid. If rotation is large enough, the subsequent flow is quasi-2D. More precisely, during the closing phase of the flaps, boundary layers of typical size $\delta=L_f/\sqrt{Re}$ form in the vicinity of each flap. At the end of the closing motion, the two boundary layers detach from the flaps and form the two vortical cores of a vortex dipole. The vorticity in a vortex core is then approximately $U_f/\delta$, and the corresponding vortex-core Rossby number is $U_f/(2 \delta \, \Omega)=Ro \sqrt{Re} $. The subsequent evolution of the vortices is strongly affected by global rotation if the vortex-core Rossby number is smaller than unity, i.e. if $Ro^2 Re < 1$.


We show in Fig.~\ref{figurms} the rms velocity fluctuations as a function of global rotation, for several flap velocities. The very first points on the left-hand-side of the figure correspond to no global rotation ($\Omega=0$). They follow the fully-3D scaling, with $u_\text{rms}$ proportional to the flap velocity (regime I), as can be seen in the inset of Fig. \ref{figurms}. All the points with nonzero rotation rate correspond to $\Omega > \nu/L_f^2=10^{-4}$ rad/s, and thus correspond to either regime II or III.

To check the scaling laws (\ref{urmsregime1}) and (\ref{urmsregime2}) for nonzero global rotation, in Fig \ref{figurmsb} we plot   $Re^{-1/2} u_{\text{rms}} / U_f$ as a function of $Ro^{-1/3} Re^{-1/6}$. This representation allows to check both laws on a single graph: it should be a linear law for regime II, whereas it should saturate to a constant value for regime III. We observe a very good collapse of the curves obtained for different values of $U_f$ onto a single master curve, with the exception of the points at $\Omega=0$ that obviously satisfy none of the scaling laws (\ref{urmsregime1}) or (\ref{urmsregime2}). Furthermore, this master curve does display a linear law for weak rotation, and it seems to saturate to a plateau for rapid rotation, as suggested by equation (\ref{urmsregime2}).

Non-rotating 3D turbulent flows at large enough Reynolds number have an rms velocity that is usually independent of the viscosity of the fluid. However, we stress the fact that regimes II and III correspond to an rms velocity that does depend on the viscosity. This is because part of the kinetic energy injected by the flaps accumulates in the large two-dimensional scales that are very weakly damped, with a dissipation rate proportional to viscosity. Although viscosity is very low, it remains a relevant parameter to evaluate the typical velocity in the present rotating turbulent flow.

\section{Which scales contain the vorticity asymmetry?\label{sec4}}

We now address the fundamental question of the scale-dependence of the cyclone-anticyclone asymmetry: as global rotation increases, which scales first develop some cyclone-anticyclone asymmetry? Is there asymmetry at every scale, or is this asymmetry contained in a few large-scale cyclones, while the small scales remain more or less symmetric? In other words, what is the typical scale of the cyclone-anticyclone asymmetry and how can it be determined experimentally? One-point statistics cannot answer this question, so in the following we consider two-point velocity correlation functions to scan the cyclone-anticyclone asymmetry at every spatial scale in the turbulent flow.

It is preferable to build two-point statistics based on velocity
instead of vorticity: as mentioned above, in most experiments based on PIV the Kolmogorov scale is not resolved. This is especially true in the asymptotic limit of turbulent flows at very large Reynolds number. A PIV vorticity field can only be accessed by approximating the velocity derivatives using finite-differences on the PIV grid. Because vorticity is dominated by small scales, different PIV resolutions would produce different values for the rms PIV vorticity. By contrast, the velocity statistics are dominated by the large scales. The velocity correlation functions are thus well measured by a PIV system even when the Kolmogorov scale is not resolved, and they allow for an unambiguous scale-dependent characterization of cyclone-anticyclone asymmetry, up to arbitrarily large Reynolds numbers.

\subsection{Antisymmetric second-order two-point velocity correlations}

Let us consider two points at positions $\textbf{x}$ and $\textbf{x}'$ in the turbulent flow. They are separated by a vector $\textbf{r}=\textbf{x}'-\textbf{x}$. We denote with a prime the quantities evaluated at point $\textbf{x}'$ and without prime the quantities evaluated at point $\textbf{x}$. We use cylindrical coordinates, where the axis $\textbf{e}_z$ is in the direction of global rotation, $\textbf{e}_r$ is the unit vector along the horizontal component of $\textbf{r}$ and $\textbf{e}_\theta$ is such that $(\textbf{e}_r,\textbf{e}_\theta,\textbf{e}_z)$ is a direct system of axes. In this coordinate system we write $\textbf{r}=r \textbf{e}_r + z \textbf{e}_z$. We further assume that, in the central region of the tank, the turbulence is homogeneous and axisymmetric, i.e. that the statistics of the turbulent velocity field are invariant to a rotation around $\textbf{e}_z$. Because in the present experiment the forcing is non-helical, we expect the bulk turbulent flow to be statistically invariant to a reflection $\mathcal{S}_\perp$ with respect to a horizontal plane. Under these assumptions, the relevant kinematics for velocity correlations are given in Ref.~\onlinecite{Lindborg}. We denote as $\textbf{R}$ the tensor of second order velocity correlations, $R_{ij}=\left< u_i(\textbf{x}) u_j(\textbf{x}') \right>=\left< u_i u'_j \right>$, where $\left< \dots \right>$ denotes statistical average. In the analysis of the PIV data, we assume ergodicity and replace this ensemble average by a time average together with a spatial average over all positions $\textbf{x}$ for which both $\textbf{x}$ and $\textbf{x}+\textbf{r}$ are inside the PIV domain.

Under the aforementioned assumptions, the general form of the tensor $\textbf{R}$ is (see for instance Ref.~\onlinecite{Lindborg})
\begin{equation}
{\bf R}=R_{rr}\, \er \er +R_{\theta \theta}\, \et \et +R_{zz}\,  \ez \ez +R_{r\theta}\,  (\er \et + \et \er)+R_{rz}\,  (\er \ez + \ez \er)+R_{\theta z}\,  (\et \ez+\ez \et) \, ,
\label{general2}
\end{equation}
where the six scalar functions $(R_{rr}, R_{\theta \theta}, \dots)$ depend on the coordinates $r$ and $z$ of the separation-vector ${\bf r}$ only. We distinguish between two kinds of velocity correlations:
\begin{itemize}
\item Symmetric correlation functions, which have an even number of $\theta$ indices (0 or 2): $R_{rr}, R_{\theta \theta}, R_{zz}$ and $R_{rz}$. They are symmetric with respect to the reflection $\mathcal{S}_\parallel$, i.e. they do not change sign under this transformation of the turbulent velocity field. These correlations are nonzero both for rotating and non-rotating turbulence.
\item Antisymmetric correlation functions, which have a single $\theta$ index: $R_{r \theta}$ and $R_{\theta z}$. They are antisymmetric with respect to the reflection $\mathcal{S}_\parallel$, i.e. they  change sign under this transformation of the turbulent velocity field. These correlations vanish for non-rotating turbulence, and they become nonzero when the $\mathcal{S}_\parallel$ symmetry is  broken by global rotation.
\end{itemize}
We focus on the antisymmetric velocity correlations, $R_{r \theta}$ and $R_{\theta z}$, because they provide a quantification of how much the turbulent velocity field breaks the $\mathcal{S}_\parallel$ symmetry, i.e. how strongly cyclone-anticyclone asymmetric this field is. We further restrict to the correlation functions that are measurable through PIV in a horizontal plane. Assuming that the turbulence is invariant to a reflexion with respect to a horizontal plane, the correlation $R_{\theta z}$ vanishes for $z=0$, that is for horizontal separation vectors. $R_{r \theta}$ is therefore the only antisymmetric second-order velocity correlation function that is nonzero for horizontal separation vectors. Lindborg and Cho\cite{LindborgCho} computed this correlation function for data of atmospheric turbulence, stressing the fact that it is very low as compared to symmetric correlation functions. Indeed, in Appendix \ref{Appendix1} we show that this correlation function vanishes identically for a two-dimensional velocity field. The measurements by Lindborg and Cho are thus compatible with their further assumption that the turbulence is quasi two-dimensional. In the present experiment, the turbulence in the central region of the arena is strongly two-dimensional for rapid global rotation. The correlation $R_{r \theta}$ is found always small: for low rotation rates, there is little cyclone-anticyclone asymmetry and $|R_{r \theta}|$ is very small. For larger rotation rates, the flow tends to two-dimensionalize and $|R_{r \theta}|$ is small as well: we measured that $|R_{r \theta}|/u^2_{\text{rms}}$ is always smaller than 0.02, for $U_f=9.2$ mm/s and all values of $\Omega$ reported in Fig.~\ref{figurms}.

\subsection{Antisymmetric third-order two-point velocity correlations: a scale-by-scale quantification of cyclone-anticyclone asymmetry}

Since no second-order velocity correlation function quantifies cyclone-anticyclone asymmetry in a quasi-two-dimensional flow, we now consider two-point third-order velocity correlation functions $S_{ijk}=\left< u_i u_j u'_k \right>$. We focus on the antisymmetric velocity correlation functions, which have an odd number of $\theta$ indices in cylindrical coordinates. In Appendix \ref{Appendix1}, we show that there are 8 such correlation functions, 3 of which can be computed from measurements of the in-plane velocity components in a horizontal PIV plane: $S_{rr\theta}, S_{\theta\theta\theta}, S_{r\theta r}$. These three correlations are in general nonzero for a velocity field that breaks the $\mathcal{S}_\parallel$ reflection symmetry, and even if this velocity field is two-dimensional. They are functions of the separation $r$, and thus provide information on the cyclone-anticyclone asymmetry of the turbulence at every scale $r$. The remaining 5 antisymmetric third-order velocity correlation functions vanish for a two-dimensional horizontal velocity field.


\begin{figure}[ht]
 \centering
 \subfigure[]{
  \includegraphics[width=65 mm]{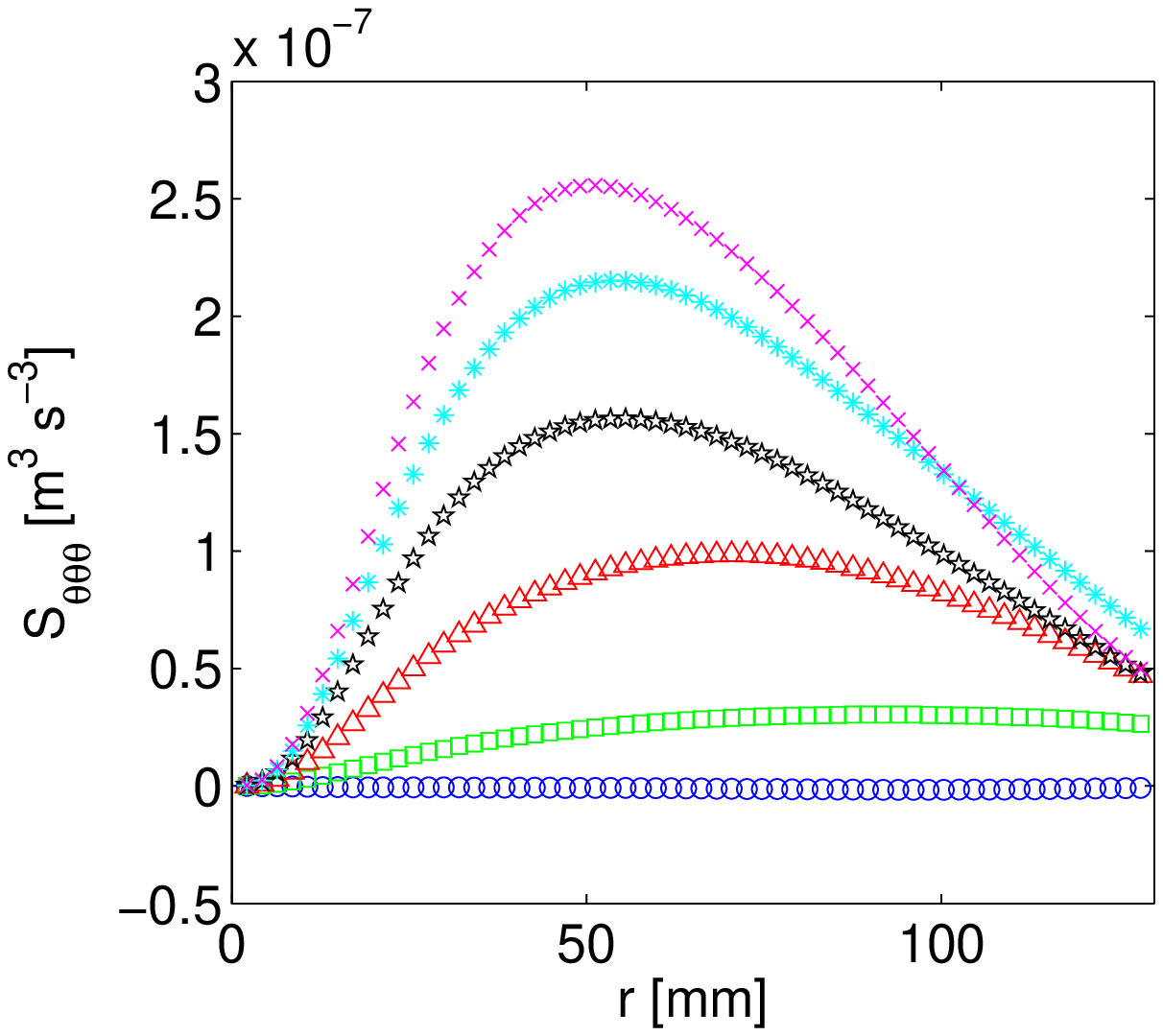}
   \label{fig:subfig1}
   }
 \subfigure[]{
  \includegraphics[width=65 mm]{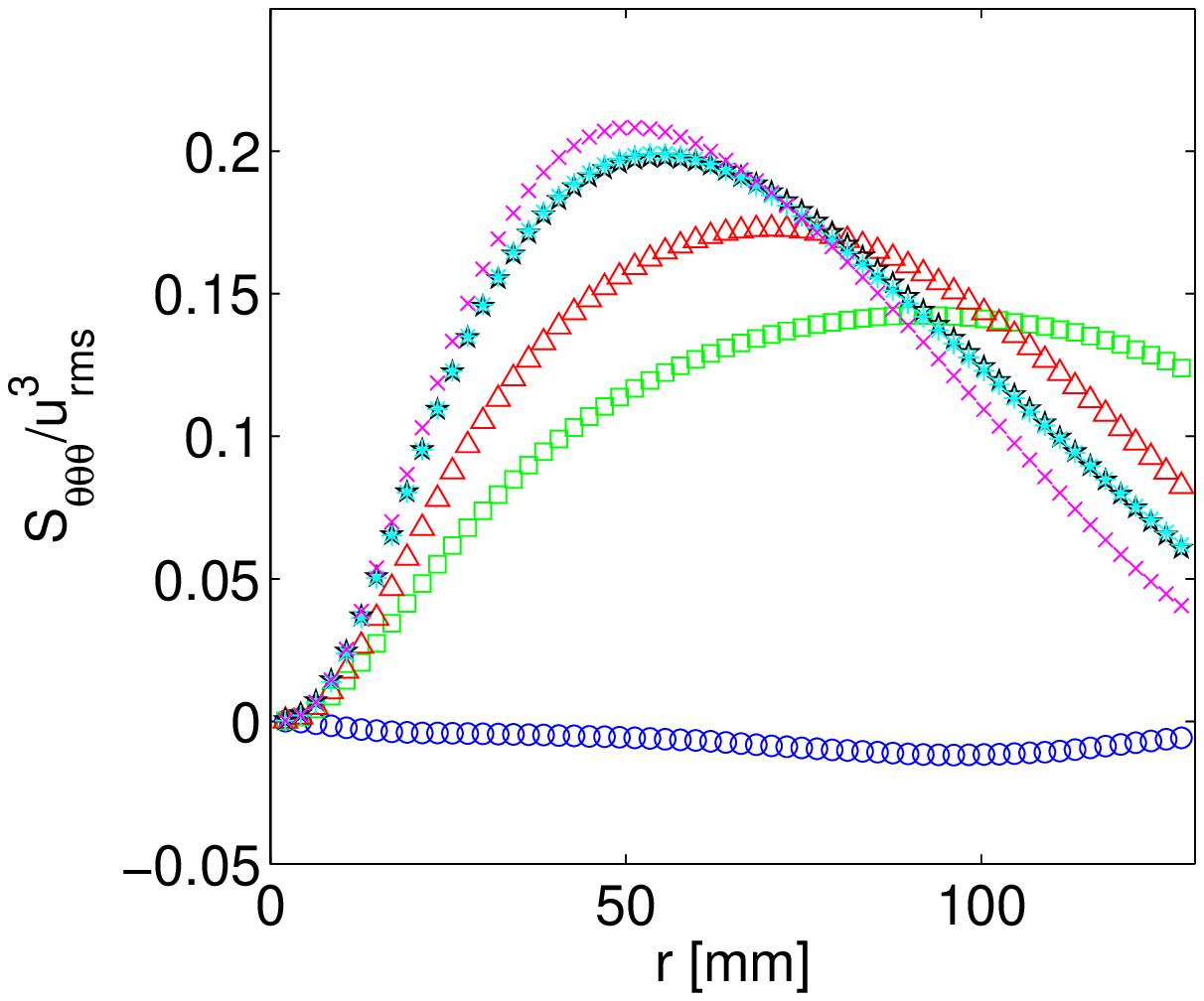}
   \label{fig:subfig2}
   }
 \subfigure[]{
  \includegraphics[width=65 mm]{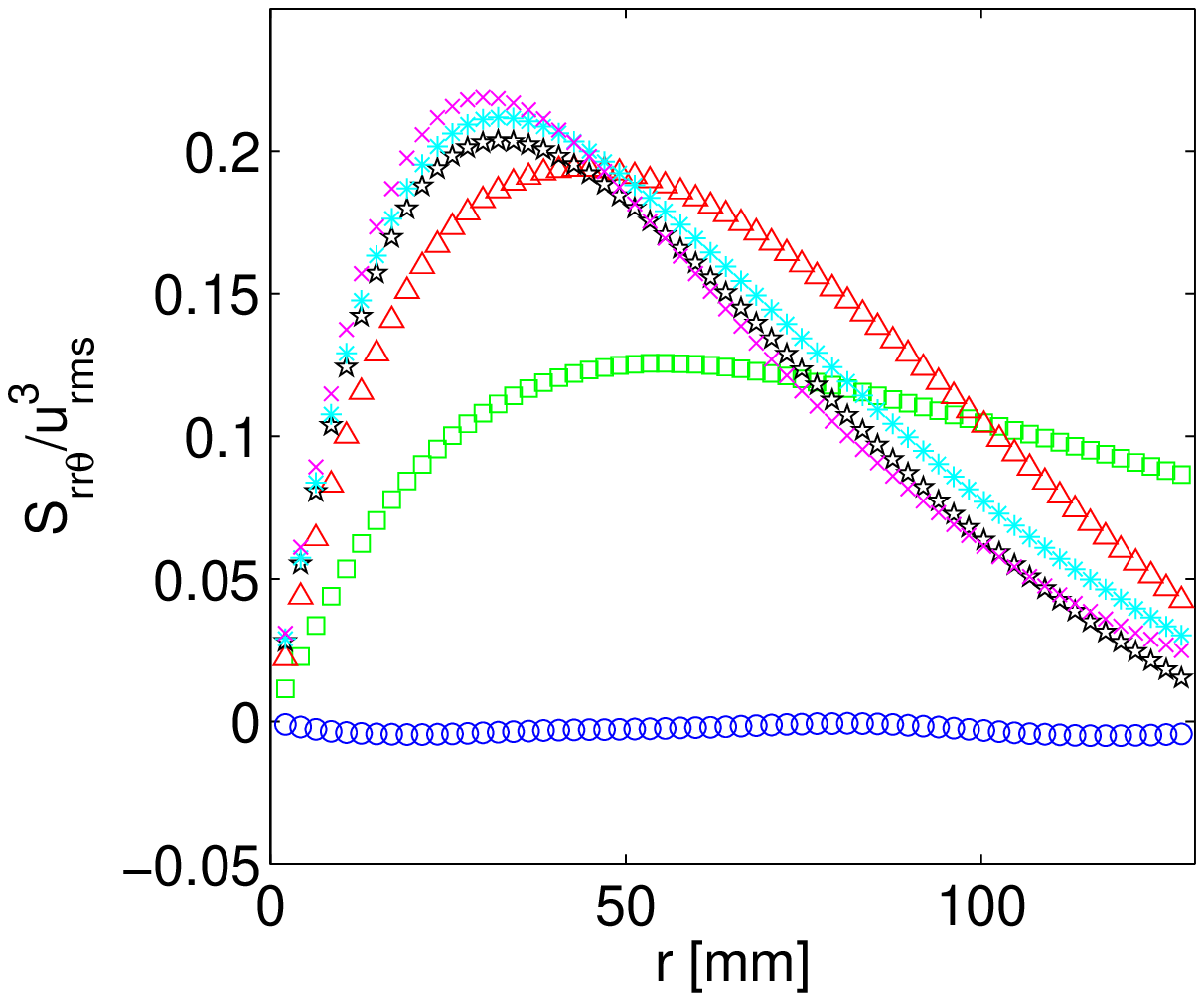}
   \label{fig:subfig3}
   }
    \subfigure[]{
  \includegraphics[width=65 mm]{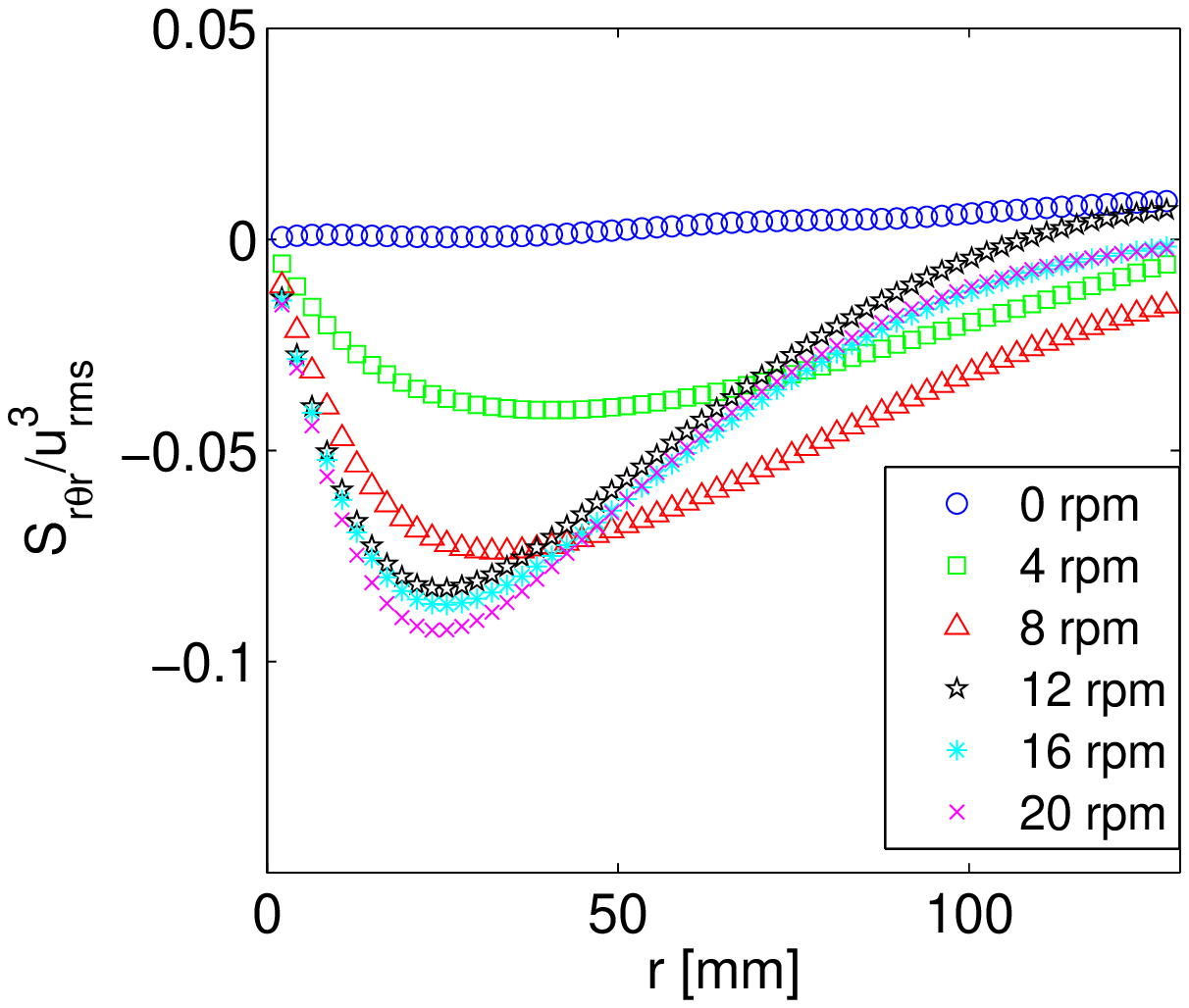}
   \label{fig:subfig3}
   }
 \caption{Antisymmetric third-order velocity correlation functions measured for $U_f=9.2$ mm/s. These correlations vanish without global rotation. (a) Correlation function $S_{\theta\theta\theta}$. The maximum value of this correlation increases with $\Omega$. (b,c,d) Correlation functions $S_{\theta\theta\theta}$,  $S_{rr\theta}$ and $S_{r\theta r}$ rescaled by $u_{\text{rms}}^3$. The three functions reach an asymptotic shape for large rotation rates. The legend in panel (d) gives the rotation rate and applies to all four panels.\label{antisymcorrel}}
\end{figure}

We plot in Fig.~\ref{antisymcorrel} the three antisymmetric third-order velocity correlation functions $S_{rr\theta}$, $S_{\theta\theta\theta}$ and $S_{r\theta r}$ for the maximum flap velocity $U_f=9.2$ mm/s and several values of $\Omega$. We observe that these correlations vanish without global rotation and become nonzero when $\Omega \neq 0$. When rescaled by $u_{\text{rms}}^3$, they reach an asymptotic shape that is independent of $\Omega$ for the largest rotation rates achieved in this experiment (see panels (b), (c) and (d)). This saturation occurs for $\Omega \geq 12$ rpm, which corresponds to the saturation of the PIV-vorticity skewness in Fig.~\ref{statvort}. 

For nonzero rotation rates, each curve in Fig.~\ref{antisymcorrel} has an extremum that indicates a characteristic scale of the cyclone-anticyclone asymmetry. Interestingly, for all three correlations the value of $r$ corresponding to the extremum is a decreasing value of $\Omega$, which seems to indicate that cyclone-anticyclone asymmetry first develops at large scales when rotation increases from zero. As $\Omega$ further increases, this characteristic scale decreases and reaches a nonzero asymptotic value: for instance, when $\Omega=20 $ rpm, $S_{\theta\theta\theta}$ has a maximum for $r \simeq 50$ mm, $S_{rr\theta}$ has a maximum for $r \simeq 30$ mm and $S_{r \theta r}$ has a minimum for $r \simeq 24$ mm. These values are comparable to the typical extension of the cores of the cyclones that we observe in snapshot \ref{snapshotsb}. 

\section{Velocity correlations for a random distribution of two-dimensional vortices\label{sec5}}

To gain intuition on antisymmetric third-order velocity correlation functions, we now compute their expressions for a field of independent two-dimensional vortices with random positions. Two-point correlations for random fields of independent vortices have been computed analytically for two-dimensional point vortices and blobs of vorticity by Chavanis and Sire\cite{Chavanis} and analytically or numerically for three-dimensional vortices.\cite{Townsend, Lundgren, Saffman, Hatakeyama, He} We would like to extend these analyses to asymmetric distributions of vortices, with an emphasis on the antisymmetric third-order correlation functions introduced in section \ref{sec4} to quantify cyclone-anticyclone 
asymmetry. This helps discuss the physics that is encoded in these correlations. The fair agreement between the analytical correlations for random vortices and the experimental correlations measured for large rotation rates indicates that the latter are dominated by the contributions from a few cyclonic vortices inside the PIV domain.

\subsection{Statistically independent random vortices}

We consider a population of $N$ identical vortices. We wish to use dimensionless variables to compute the velocity correlation functions: we write the vorticity field of one vortex as $\omega_0 W(|{\bf x}-{\bf x_0}|/r_0)$, where ${\bf x_0}$ is the position of the vortex center,  $\omega_0$ is the typical vorticity of the vortex, $r_0$ is the extent of the vortex core, and $W(s)$ is the dimensionless vorticity profile. The corresponding azimuthal velocity profile of the vortex is $r_0 \omega_0 V(|{\bf x}-{\bf x_0}|/r_0)$, where the dimensionless velocity profile $V(s)$ is related to $W(s)$ by $W(s)= s^{-1} d (s V(s)) / ds$.
In the following we non-dimensionalize the lengths with the scale $r_0$, the velocities with $r_0 \omega_0$ and the vorticities with $\omega_0$. We consider that the $N$ vortices are in a square domain of dimensional area $ Lr_0 \times L r_0$, where $L$ is the dimensionless side length of the square. The dimensional vortex density is $\tilde{n}=N/(r_0 L)^2$ and its dimensionless counterpart is $n=N/L^2$.

It is important to consider vortices with vanishing circulation at large distance from the vortex center: vortices with nonzero circulation would lead to a logarithmic divergence of the kinetic energy with the domain size, whereas we expect the motion of a flap to provide a finite amount of kinetic energy to the neighboring fluid in a region of typical extension $L_f$ close to the flap. We thus consider shielded vortices with dimensionless profile
\begin{equation}
W(s)=e^{-s^2}-a^2 e^{-a^2 s^2}\, , \, \qquad V(s)=\frac{e^{-a^2 s^2}-e^{-s^2}}{2s}\, ,
\label{profile2}
\end{equation}
where $a$ is a dimensionless shielding parameter. We further consider $0<a\ll1$: such a shielded-vortex has a core of extension $s \sim 1$, and a maximum vorticity $W \simeq 1$ for $s=0$. It resembles a Gaussian vortex profile with positive vorticity for $0<s \ll 1/a$. However, the velocity and vorticity of the vortex rapidly decay to zero above the ``shielding length" $1/a$, so that the vortex has vanishing circulation at large distance from its center. In dimensional units, the vortex has a maximum vorticity of approximately $\omega_0$ at its center, a core of extension $r_0$, and its velocity and vorticity decay rapidly above the dimensional shielding length $r_0/a$.

From now on and until section \ref{comparison}, we use dimensionless positions, velocity and vorticity, which we write similarly to their dimensional counterparts to alleviate notations. Denoting the (dimensionless) positions of the vortex centers as $(\textbf{x}_i)_{i \in \{1,N\}}$, the vorticity of the random field is
\begin{equation}
\omega(\textbf{x})=\displaystyle\sum_{i=1}^{N} W(|{\bf x}-{\bf x_i}|)\, .
\end{equation}
We consider independent vortices, whose positions $({\textbf{x}_i})_{i \in  \{1,N\}}$ are independent random variables uniformly distributed in the domain of size $L\times L$. Statistical average $\left< \hdots \right>$ then corresponds simply to an integration with respect to all the positions $({\bf x_i})_{i \in  \{1,N\}}$, and division by $L^{2N}$. For instance, the average vorticity field is
\begin{equation}
\left< \omega(\textbf{x}) \right>=\frac{1}{L^{2N}} \int \displaystyle \left( \displaystyle\sum_{j=1}^{N} W(|{\bf x}-{\bf x_j}|) \right) \prod_{i=1}^{N} \mathrm{d} {\bf x_i}  = \frac{N}{L^{2}} \int  W(|{\bf x}-{\bf x_1}|) \mathrm{d} {\bf x_1}\, ,
\end{equation}
where the integrations are performed on the domain of size $L\times L$. We focus on the infinite-domain limit: $L \to \infty$, $N \to \infty$, the vortex density $n=N/L^2$ being held constant. Let us denote this limit as $\displaystyle\lim_{(L, N) \to \infty}$. Using profile (\ref{profile2}) we obtain $\left< \omega(\textbf{x}) \right>=0$: on statistical average, these fields of randomly distributed vortices have vanishing vorticity and velocity.

\subsection{Third-order antisymmetric correlation functions}

\begin{figure}
\includegraphics[width=120 mm]{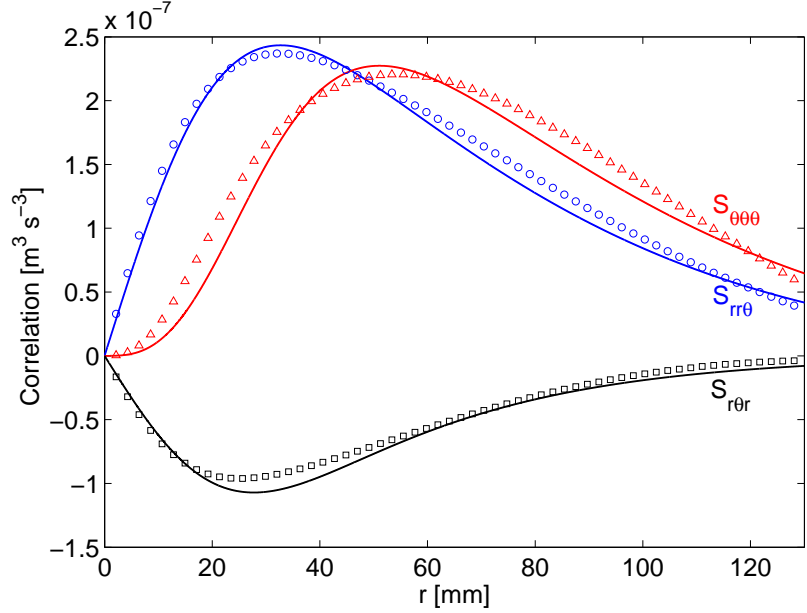}
\caption{Comparison between the third-order antisymmetric velocity correlation functions measured in the experiment and computed for the model of randomly distributed independent vortices. Symbols are correlations measured in the experiment for rapid global rotation ($U_f=9.2$ mm/s and $\Omega=16$ rpm). Solid lines are from the model, using a shielding parameter $a=0.15$, a vortex size $r_0=17$ mm and a prefactor $\tilde{n} r_0^5 \omega_0^3=2.65\, 10^{-6}$ m$^{3}$s$^{-3}$.}
\label{Ordre3}
\end{figure}

We now compute the antisymmetric third-order velocity correlation functions that characterize cyclone-anticyclone asymmetry in the fields of randomly distributed vortices. To take advantage of the relatively simple Gaussian shape of the vorticity profile, we first compute the velocity-vorticity correlations $\left<  (u_r^2+u_\theta^2) \omega' \right>(r)$ and $\left<  (u_r^2-u_\theta^2) \omega' \right>(r)$ before going back to these velocity correlations. Denoting as $u_x$ and $u_y$ the two horizontal velocity components along the $x$ and $y$ Cartesian axes, the first correlation reads in the infinite-domain limit
\begin{eqnarray}
& & \left< (u_r^2+u_\theta^2)   \omega' \right>  = \left< (u_x^2+u_y^2)   \omega(\textbf{x'}) \right>  \\
\nonumber & &  =  \displaystyle\lim_{(L, N) \to \infty}  \left< \left[\left( \displaystyle\sum_{i=1}^{N} \frac{y-y_i}{| {\bf x}-{\bf x_i} |} V(|{\bf x}-{\bf x_i}|)     \right)^2 + \left( \displaystyle\sum_{i=1}^{N} \frac{x-x_i}{| {\bf x}-{\bf x_i} |} V(|{\bf x}-{\bf x_i}|)     \right)^2\right] \left( \displaystyle\sum_{i=1}^{N} W(|{\bf x'}-{\bf x_i}|) \right) \right> \label{eqstat}\\
\nonumber & &  =  \displaystyle\lim_{(L, N) \to \infty} \displaystyle\sum_{i=1}^{N} \left<  V(|{\bf x}-{\bf x_i}|)^2\,  W(|{\bf x'}-{\bf x_i}|)    \right>  \\
\nonumber & &  =  \displaystyle\lim_{(L, N) \to \infty}  N \left<  V(|{\bf x}-{\bf x_1}|)^2\,  W(|{\bf x'}-{\bf x_1}|)    \right> = \displaystyle\lim_{(L, N) \to \infty} N \left<  V(|{\bf x_1}|)^2\,  W(|{\bf x_1}-{\bf r}|)    \right>  \,. 
\end{eqnarray}
Because of axisymmetry, this correlation is independent of the direction of the (horizontal) separation vector $\textbf{r}$. Upon choosing $\textbf{r}=r\, \textbf{e}_x$ and denoting as $x_1$ and $y_1$ the coordinates of $\textbf{x}_1$, we obtain
\begin{eqnarray}
& & \left< (u_r^2+u_\theta^2)   \omega' \right>  =  n \int_{x_1=-\infty}^{x_1=+\infty}  \int_{y_1=-\infty}^{y_1=+\infty} \left[V\left(\sqrt{x_1^2+y_1^2}\right)\right]^2 W\left(\sqrt{(x_1-r)^2+y_1^2}\right) \mathrm{d}x_1 \mathrm{d}y_1 \qquad  \\
\nonumber & &  =   -\frac{\pi n}{4} \left[  e^{- r^2} \left(2 \text{Ei}\left(\frac{r^2}{a^2+2}\right)-\text{Ei}\left(\frac{r^2}{2 a^2+1}\right)-\text{Ei}\left(\frac{r^2}{3}\right)\right) \right. \\
\nonumber & &+  \left.  a^2 e^{-a^2 r^2} \left(\text{Ei}\left(\frac{a^2 r^2}{3}\right)+\text{Ei}\left(\frac{a^4 r^2}{a^2+2}\right)-2 \text{Ei}\left(\frac{a^4 r^2}{2 a^2+1}\right)\right)\right]\, .
\end{eqnarray}
In this equation, the exponential integral function is defined by 
\begin{equation}
\Ei{x}=\int \limits_{-\infty}^x \frac{e^t}{t} \mathrm{d}t\, ,
\end{equation}
where the principal value of the integral is taken when the argument $x$ is positive.

Using axisymmetry, the velocity-vorticity correlation $\left<  (u_r^2-u_\theta^2) \omega' \right>(r)$ reads
\begin{eqnarray}
\left< (u_r^2-u_\theta^2)   \omega' \right> & = & \left< \left[ u_x^2(\textbf{x}) -u_y^2(\textbf{x})  \right]   \omega(\textbf{x}+r \textbf{e}_x) \right> \\
\nonumber & = & \displaystyle\lim_{(L,N) \to \infty} N \left< \frac{y_1^2-x_1^2}{x_1^2+y_1^2} \left[V\left(\sqrt{x_1^2+y_1^2}\right)\right]^2   W\left(\sqrt{(x_1-r)^2+y_1^2}\right) \right> \\
\nonumber & = & n \int_{x_1=-\infty}^{x_1=+\infty}  \int_{y_1=-\infty}^{y_1=+\infty}  \frac{y_1^2-x_1^2}{x_1^2+y_1^2} \left[V\left(\sqrt{x_1^2+y_1^2}\right)\right]^2   W\left(\sqrt{(x_1-r)^2+y_1^2}\right) \mathrm{d}x_1 \mathrm{d}y_1 \\
\nonumber & = & \frac{\pi n}{4 r^2}  \left[ 2 \left(a^2+2\right) e^{-\frac{a^2+1}{a^2+2} r^2}+\left(\frac{2}{a^2}+1\right) e^{-\frac{2 a^2 }{a ^2+2} r^2}-\left(2 a^2+1\right) e^{-\frac{2 a^2 }{2 a^2+1} r^2} \right. \\
\nonumber & & \left.  +  \left(-\frac{2}{a^2}-4\right) e^{-\frac{a^2 \left(a^2+1\right)}{2 a^2+1} r^2}+3 e^{\frac{-2 a^2}{3} r^2}-3 e^{-\frac{2}{3} r^2} \right] \,.
\end{eqnarray}

%
%

In Appendix \ref{Appendix2}, we relate these velocity-vorticity correlations to the velocity correlations $S_{\theta \theta \theta}$, $S_{r r \theta}$, $S_{r \theta r}$ for an axisymmetric, incompressible and homogeneous two-dimensional velocity field. The system of equations to solve is
\begin{eqnarray}
\mathrm{d}_r (r S_1) & = & r \left<  (u_r^2+u_\theta^2 ) \omega' \right> \mbox{, with } S_1=S_{r r \theta}+S_{\theta \theta \theta} \label{eqS1}\\
\mathrm{d}_r (r S_2)-2S_2  & = & r \left<  (u_r^2-u_\theta^2) \omega' \right> \mbox{, with } S_2=S_{r r \theta}-S_{\theta \theta \theta}-2 S_{r \theta r} \label{eqS2}\\
\mathrm{d}_r (r S_3)+2S_3  & = & r \left<  (u_r^2-u_\theta^2) \omega' \right> \mbox{, with } S_3=S_{r r \theta}-S_{\theta \theta \theta}+2 S_{r \theta r} \label{eqS3}
\end{eqnarray}
We solve this system for $S_1$, $S_2$ and $S_3$ before going back to the original correlations. The velocity correlations $S_{\theta \theta \theta}$, $S_{r r \theta}$, $S_{r \theta r}$ are thus given by
\begin{eqnarray}
S_{rr\theta} & = & \frac{1}{4} (2S_1+S_2+S_3) \, \label{SfS1} ,\\
S_{\theta \theta\theta} & = & \frac{1}{4} (2S_1-S_2-S_3) \, , \label{SfS2} \\
S_{r\theta r} & = & \frac{1}{4} (S_3-S_2) \, \label{SfS3} ,
\end{eqnarray}
where $S_1$, $S_2$ and $S_3$ are the solutions of equations (\ref{eqS1} - \ref{eqS3}) that vanish for $r \to 0$ and $r \to \infty$:
\begin{eqnarray}
S_1(r) & = & \frac{\pi n}{8 r} \left[  \Ei{-\frac{2r^2}{3}} -\Ei{-\frac{2 a^2 r^2}{3}} -\Ei{-\frac{2 a^2 r^2}{2+a^2}} -2 \Ei{-\frac{1+a^2}{2+a^2}r^2}   \right. \\
\nonumber & & +2  \Ei{-\frac{a^2(1+a^2)}{1+2a^2}r^2} + \left.  \Ei{-\frac{-2a^2}{1+2a^2}r^2} \right] + e^{-r^2} \left[ -\Ei{\frac{r^2}{3}} +2 \Ei{\frac{r^2}{2+a^2}} \right. \\
\nonumber & & \left.  - \Ei{\frac{r^2}{1+2a^2}} \right] +  e^{-a^2 r^2} \left[ \Ei{\frac{a^2 r^2}{3}} + \Ei{\frac{a^4 r^2}{2+a^2}} - 2 \Ei{\frac{a^4 r^2}{1+2a^2}} \right] \,  , \\
S_2(r) & = & -\frac{\pi n}{8 r} \left[ -3 e^{-\frac{2}{3} r^2}+ 3 e^{-\frac{2}{3} a^2 r^2}+\left( 1+\frac{2}{a^2} \right) e^{-\frac{2a^2}{2+a^2}r^2}+(4+2a^2) e^{-\frac{1+a^2}{2+a^2}r^2}-(1+2a^2) e^{-\frac{2a^2}{1+2a^2}r^2} \right. \qquad \\
\nonumber & & - \left. \left( 4+\frac{2}{a^2} \right) e^{-\frac{a^2+a^4}{1+2a^2}r^2} \right] +\frac{\pi n} {4} r \left[  \Ei{-\frac{2r^2}{3}} -a^2 \Ei{-\frac{2a^2 r^2}{3}} - \Ei{-\frac{2 a^2 r^2}{2+a^2}} \right. \\
\nonumber & &  -(a^2+1) \Ei{-\frac{(1+a^2)r^2}{2+a^2}} +  \left. a^2 \Ei{-\frac{2 a^2 r^2}{1+2a^2}}+(a^2+1) \Ei{-\frac{a^2(a^2+1)r^2}{1+3a^2}} \right] \, , \\
S_3(r) & = & -\frac{\pi n}{16 a^2 r^3} \left[ -9 a^2 e^{-\frac{2}{3} r^2}+ 9 e^{-\frac{2}{3} a^2 r^2}+\frac{(2+a^2)^2}{a^2} e^{-\frac{2a^2}{2+a^2}r^2}+4a^2 \frac{(2+a^2)^2}{1+a^2} e^{-\frac{1+a^2}{2+a^2}r^2} \right. \\
\nonumber & & -(1+2a^2)^2 e^{-\frac{2a^2}{1+2a^2}r^2} -  \left. 4 \frac{(1+2a^2)^2}{a^2(1+a^2)}  e^{-\frac{a^2(1+a^2)}{1+2a^2}r^2} \right] \, .
\end{eqnarray}

%
%

\subsection{Comparison with experiments for rapid global rotation \label{comparison}}


We compare in Fig.~\ref{Ordre3} the third-order antisymmetric velocity correlation functions measured in the experiment for $\Omega = 16$~rpm and the model (\ref{SfS1}-\ref{SfS3}) of randomly distributed independent vortices. The dimensional velocity correlations are $r_0^3 \omega_0^3 S (r/r_0)$, where $S$ is one of the three dimensionless functions given by (\ref{SfS1}-\ref{SfS3}). They depend upon three fitting parameters: the extension $r_0$ of the vortex core, the shielding parameter $a$, and a prefactor $\tilde{n} r_0^5 \omega_0^3 = n r_0^3 \omega_0^3 $. In Fig.~\ref{Ordre3} we use the values $r_0=17$ mm, $a=0.15$ and  $\tilde{n} r_0^5 \omega_0^3=2.65\, 10^{-6}$ m$^{3}$s$^{-3}$. 

Despite their intricate analytical expressions, the shapes of the model correlation functions are quite simple and reproduce well the correlations measured in the experiment for rapid rotation. They present an extremum for a finite $r$ that scales like the vortex core radius, before decaying to zero for large $r$. Close to zero, both $S_{rr\theta}$ and $S_{r\theta r}$ start off with a linear dependence in $r$, whereas $S_{\theta \theta \theta}$ grows as $r^3$. 
In Appendix \ref{Appendix3}, we relate the behavior of these correlations for small separation $r$ to one-point statistics of the velocity gradients, therefore showing the link between the antisymmetric velocity correlation functions and the vorticity asymmetry $\left< \omega^3 \right>$. The former appears as a natural scale-dependent extension of the latter.

The values of the fitting parameters are compatible with the snapshot \ref{snapshotsb}. The value of $r_0$ used in Fig.~\ref{Ordre3} is comparable to the typical radius of the cyclones in this figure. Assuming that the motion of the flaps sets into motion the fluid in a region of horizontal extension $L_f$, the shielding parameter should be roughly $a \sim r_0/L_f = 0.17$, which is compatible with the value $a=0.15$ used in Fig.~\ref{Ordre3}. 

The value of $\tilde{n} r_0^5 \omega_0^3$ is difficult to evaluate in the experiment, because it requires a precise definition of a cyclone, as opposed to any local maximum in the vorticity field. By visual inspection, we observe typically $N \simeq 4$ cyclones in the PIV domain with central vorticity between $3$ s$^{-1}$ and $8$ s$^{-1}$. The corresponding $\tilde{n} r_0^5 \omega_0^3$ is of the order of $10^{-5}$ m$^{3}$s$^{-3}$, which is larger than the fitting value by a factor of four. 
Because an error of only 30\% on $r_0$ can explain this factor of four, we consider that the agreement is still satisfactory. Another origin of this discrepancy may be the fact that there are still a few anticyclones in the PIV domain: in the model we consider a single population of identical cyclones with surface density $\tilde{n}$, and this $\tilde{n}$ should correspond roughly to the difference between the cyclone and anticyclone surface densities in the experiment. The effective $\tilde{n}$ that gives a good agreement between the model and the data is thus smaller than the cyclone density in the experiment.

Because a set of randomly distributed identical vortices is an oversimplified model of a turbulent flow with vortices of various amplitudes and velocity profiles, we do not go into further quantitative comparison between the model and the experimental data. Nevertheless, it is remarkable that this model captures the shapes of the experimental antisymmetric correlation functions, despite its simplicity. This indicates that, in this experiment and for large rotation rates, cyclone-anticyclone asymmetry  is dominated by a few large-scale cyclones.

\section{Conclusion\label{sec6}}


We performed forced rotating turbulence experiments in a statistically steady state. We first characterized the evolution of the rms velocity with the rotation rate. The measured velocities are compatible with the following phenomenology: in the rapid rotation regime, part of the kinetic energy injected by the flaps accumulates in quasi-2D motions which are weakly damped as compared to 3D motions, with an energy dissipation rate proportional to viscosity. As a result, the rms velocity is larger in the rapidly rotating regime than in the non-rotating one.

Because global rotation imposes a preferred direction of rotation, the flow breaks the cyclone-anticyclone symmetry. We introduced a set of antisymmetric third-order velocity correlation functions which provide a scale-by-scale quantification of this asymmetry. We also computed the analytical expression of these correlation functions for a field of independent random cyclones. The good qualitative agreement between the experimental correlation functions and this simple model indicates that much of the cyclone-anticyclone asymmetry originates from a few large-scale long-lived cyclones in the experiment. 

These antisymmetric velocity correlations therefore appear as natural tools to quantify cyclone-anticyclone asymmetry in rotating turbulent flows. In the present experiment, the Reynolds number is moderate and we used these correlations essentially to identify the dominant scale of the cyclone-anticyclone asymmetry. Nevertheless, the question of their possible universal behavior for rotating flows with very large Reynolds number naturally arises. Indeed, in homogeneous and isotropic turbulence, symmetric velocity correlation functions  satisfy the celebrated von K\'arm\'an-Howarth relation. Are there equivalent relations for antisymmetric velocity correlation functions in the inertial range of fully-developed rotating turbulence? How do these correlations depend on the Rossby number? The atmospheric data analysis by Lindborg and Cho\cite{LindborgCho} indicates an $r^2$ dependence for antisymmetric structure functions, but the origin of this scaling regime remains to be understood. No such power-law regimes were observed in the present experiment, because of the moderate values of the Reynolds number.
It would be interesting to apply the present statistical tools to numerical simulations of homogeneous rotating turbulence, in which the spatial inhomogeneities inherent to the experimental forcing could be avoided.

Finally, one could investigate the behavior of these  antisymmetric velocity correlation functions in helical turbulence and in other systems of geophysical relevance, such as rotating turbulence in the presence of stratification and/or magnetic field, to get some insight in the scale-by-scale cyclone-anticyclone asymmetry of astrophysical, oceanic and atmospheric flows (note that for  turbulence lacking reflection-symmetry with respect to horizontal planes, a more general form of correlation tensors should be considered, as explained in Ref.~\onlinecite{Oughton}). Because they are based on two-point velocity measurements only, these tools are particularly well-suited to quantify cyclone-anticyclone asymmetry in oceanographic and atmospheric data, which typically correspond to large-Reynolds-number flows that are either sampled on a rather coarse spatial grid or measured by a single moving probe.

We acknowledge P. Augier, P. Billant, J.-M. Chomaz for kindly providing
the flap apparatus, and A. Aubertin, L. Auffray, C. Borget and R. Pidoux
for their experimental help. FM would like to thank the Institut Universitaire de France for its support. This work is supported by the ANR grant no.
2011-BS04-006-01 ``ONLITUR''. The rotating platform ``Gyroflow''
was funded by the ``Triangle de la Physique''.


\appendix

\section{Two-point correlations for homogeneous, axisymmetric and $\mathcal{S}_\perp$-symmetric turbulence \label{Appendix1}}

\subsection{Second-order correlations}
Because of homogeneity, the tensor of second-order velocity correlations is symmetric. Using polar coordinates and assuming axisymmetry (invariance to rotations around the vertical axis) and invariance to $\mathcal{S}_\perp$ (reflection-symmetry with respect to horizontal planes), it is given by equation (\ref{general2}), where the scalar functions $R_{ij}$ depend on the coordinates $r$ and $z$ of the separation-vector ${\bf r}$, but not on polar angle $\theta$. However, the unit vectors $\er$ and $\et$ do depend on $\theta$, with $\partial_\theta \er=\et$ and $\partial_\theta \et=-\er$. Because of incompressibility, the tensor ${\bf R}$ satisfies ${\boldsymbol{\nabla}} \cdot_1 {\bf R} = {\bf 0}$, where $\cdot_1$ means contraction with respect to the first index. Using the polar expression for operator ${\boldsymbol{\nabla}}$, we obtain
\begin{eqnarray}
{\bf 0} & = & {\boldsymbol{\nabla}} \cdot_1 {\bf R} = \left( {\er \partial_r + \et \frac{\partial_\theta}{r} +\ez \partial_z }\right) \cdot_1 \left[ R_{rr}\, \er \er +R_{\theta \theta}\, \et \et  +R_{zz}\,  \ez \ez \right. \\
\nonumber & + & \left.    R_{r\theta}\,  (\er \et + \et \er)+R_{rz}\,  (\er \ez + \ez \er)+R_{\theta z}\,  (\et \ez+\ez \et) \right] \, .
\end{eqnarray}
Performing the derivatives and the contraction, we obtain a vector quantity whose three components must vanish. This gives three relations between the components of tensor ${\bf R}$:
\begin{eqnarray}
0 & = & \frac{1}{r} \partial_r (r R_{rr})-\frac{R_{\theta \theta}}{r}+\partial_z R_{rz} \label{div2er} \, ,\\
0 & = & \frac{1}{r^2} \partial_r (r^2 R_{r \theta})+\partial_z R_{\theta z} \, \label{divanti} ,\\
0 & = & \frac{1}{r} \partial_r (r R_{rz})+\partial_z R_{zz} \, .
\end{eqnarray}
Out of these three equations, only (\ref{divanti}) is a constraint on the antisymmetric correlation functions. In the particular case of a two-dimensional velocity field, $\partial_z R_{\theta z} =0$ and the constraint becomes that $r^2 R_{r \theta}$ is a constant. This constant must be zero to avoid any divergence of $R_{r \theta}$, hence $R_{r \theta}$ vanishes.

\subsection{Third-order correlations}

The third-order velocity correlation tensor ${\bf S}$ is symmetric with respect to its first two indices and can be expressed in terms of 18 scalar functions (see for instance Ref. \onlinecite{Lindborg})
\begin{eqnarray}
& & {\bf S}  =   S_{rrr}\, \er \er \er + S_{rr \theta} \, \er \er \et +S_{\theta \theta r}\, \et \et \er +S_{r \theta r}\, (\er \et +\et \er)\er + S_{zz r} \, \ez \ez \er \label{generalS}\\
\nonumber & & +S_{\theta \theta \theta}\, \et \et \et + S_{zzz}\, \ez \ez \ez + S_{rrz}\, \er \er \ez + S_{\theta\theta z}\, \et \et \ez + S_{rzz}\, (\er \ez + \ez \er) \ez  \\
\nonumber & & + S_{rzr}\, (\er \ez + \ez \er) \er +S_{\theta z \theta}\, (\et \ez + \ez \et) \et + S_{zz\theta}\, \ez \ez \et +S_{\theta z z}\, (\et \ez + \ez \et) \ez   \\
\nonumber & & + S_{\theta z r}\, (\et \ez + \ez \et) \er + S_{r \theta z}\, (\er \et + \et \er) \ez +S_{r z \theta}\, (\er \ez + \ez \er) \et + S_{r \theta \theta}\, (\er \et + \et \er) \et  \, ,
\end{eqnarray}
where the scalar functions depend upon $r$ only for a two-dimensional axisymmetric and homogeneous velocity field. The antisymmetric third-order velocity correlation functions are the $S_{ijk}$ with an odd number of $\theta$ indices. There are 8 such functions. Because of $\mathcal{S}_\perp$ symmetry (non-helical turbulence), the correlations with an odd number of $z$ indices vanish when the separation vector is horizontal, i.e. for $z=0$. There are thus 5 antisymmetric third-order correlation functions that do not vanish when measured in a horizontal plane: $S_{rrr}$, $S_{rr \theta}$, $S_{r \theta r}$, $S_{zz\theta}$ and $S_{\theta z z}$. Since we measure only the horizontal velocity components, we can determine three of these correlations: $S_{rrr}$, $S_{rr \theta}$, $S_{r \theta r}$. Note that for a two-dimensional velocity field, the vertical velocity is zero, so $S_{zz\theta}$ and $S_{\theta z z}$ vanish.

\section{From velocity-vorticity correlations to third-order velocity correlations  \label{Appendix2}}

Gaussian vorticity profiles make the velocity-vorticity correlations easier to compute than the velocity correlations. Here we derive the link between these correlations and the third-order antisymmetric velocity correlations that we measure experimentally. Using Cartesian coordinates, the tensor $\left<u_i u_j \omega' \right>$ is the curl with respect to ${\bf r}$ of the third-order velocity correlation tensor $S_{ijk}=\left<u_i u_j u'_k \right>$. We wish to compute this curl in polar coordinates. We now focus on two-dimensional horizontal velocity fields only, so that any scalar correlation function in (\ref{generalS}) with a $z$ index is set to zero:
\begin{eqnarray}
{\bf S} & = & S_{rrr}\, \er \er \er + S_{rr \theta} \, \er \er \et +S_{\theta \theta r}\, \et \et \er \\
\nonumber & + & S_{r \theta r}\, (\er \et +\et \er)\er + S_{r \theta \theta}\, (\er \et + \et \er) \et +S_{\theta \theta \theta}\, \et \et \et \, .
\end{eqnarray}
The gradient of ${\bf S}$ is a fourth-order tensor that reads
\begin{equation}
 {\boldsymbol{\nabla}} {\bf S} =  \left( \er \partial_r + \et \frac{\partial_\theta}{r} \right)     {\bf S} \label{gradS} \, .
\end{equation}
Let us first make use of incompressibility: the tensor ${\bf S}$ is divergence free with respect to its third index, that is ${\boldsymbol{\nabla}} \cdot_3 {\bf S}={\bf 0}$. The quantity ${\boldsymbol{\nabla}} \cdot_3 {\bf S}$ is also the contraction of (\ref{gradS}) over indices 1 and 4. The resulting second-order tensor must vanish, which gives three constraints:
\begin{eqnarray}
 \mathrm{d}_r (r S_{rrr})-2 S_{r \theta \theta} & = & 0 \, ,\\
 \mathrm{d}_r (r S_{\theta \theta r})+2 S_{r \theta \theta} & = & 0  \, ,\\
\mathrm{d}_r (r S_{r\theta r})+S_{rr\theta}-S_{\theta \theta \theta} & = & 0 \, \label{contrainte3} .
\end{eqnarray}
The present study deals with antisymmetric third-order correlation functions, so we use constraint (\ref{contrainte3}) only in what follows.

The velocity-vorticity correlations $\left<u_i u_j \omega' \right>$ are obtained by taking the product of ${\boldsymbol{\nabla}} {\bf S}$ along indices 1 and 4 with the Levi-Civita antisymmetric tensor $\epsilon_{ij}$. For this two-dimensional velocity field, vorticity is along the vertical only and the result is
\begin{eqnarray}
[ {\boldsymbol{\nabla}} {\bf S}]_{r..\theta} -  [{\boldsymbol{\nabla}} {\bf S}]_{\theta.. r} & = & \left<  u_r^2 \omega' \right> \er \er +\left<  u_\theta^2 \omega' \right> \et \et + \left<  u_r u_\theta \omega' \right> (\er \et +\et \er) \\
\nonumber & = & \left[ \frac{1}{r} \mathrm{d}_r (r S_{rr\theta}) + \frac{2}{r} S_{r\theta r} \right] \er \er +  \left[ \frac{1}{r} \mathrm{d}_r (r S_{\theta \theta \theta}) - \frac{2}{r} S_{r\theta r} \right] \et \et \\
\nonumber & + &  \left[ \frac{1}{r} \mathrm{d}_r (r S_{r \theta\theta}) - \frac{1}{r} S_{r r r} +  \frac{1}{r} S_{\theta \theta r}\right] (\er \et + \et \er)
\end{eqnarray}
The sum of the $\er \er$ and $\et \et$ components leads to equation (\ref{eqS1}). The difference yields
\begin{equation}
\mathrm{d}_r [r (S_{r r\theta } - S_{\theta \theta\theta})]+4 S_{r\theta r}=r  \left<  (u_r^2-u_\theta^2) \omega' \right>  \label{eqdiff}
\end{equation}
 The linear combinations (\ref{eqdiff})-2$\times$(\ref{contrainte3}) and  (\ref{eqdiff})+2$\times$(\ref{contrainte3}) are equations (\ref{eqS2}) and (\ref{eqS3}).

\section{Expansion of the antisymmetric correlation functions for small separation $r$ \label{Appendix3}}

Let us assume that the turbulence has homogeneous and axisymmetric statistics and start off by expanding $S_{\theta \theta \theta}(r)$ for small $r$. 
Using Cartesian coordinates $(x,y)$ in the horizontal:
\begin{eqnarray}
S_{\theta \theta \theta}(r) & = & \la u_y^2(\textbf{0}) u_y(r \textbf{e}_x) \ra \\
\nonumber & = & \la u_y^2|_{\textbf{0}} \left( u_y|_{\textbf{0}} + r\partial_x(u_y)|_{\textbf{0}} +\frac{r^2}{2}\partial_{xx}(u_y)|_{\textbf{0}}+\frac{r^3}{6}\partial_{xxx}(u_y)|_{\textbf{0}} +\frac{r^4}{24}\partial_{xxxx}(u_y)|_{\textbf{0}} \right) \ra + \mathcal{O}(r^5) \, , \\
\end{eqnarray}
where the subscript $|_{\textbf{0}}$ means ``evaluated at $\textbf{x}=\textbf{0}$". The term proportional to $r^0$ vanishes because of axisymmetry. The coefficient of the term proportional to $r$ is $\la u_y^2\partial_x u_y \ra=\partial_x \la u_y^3 \ra /3 = 0$, because of homogeneity. The $r^2$ coefficient is proportional to $\la u_y^2 \partial_{xx}(u_y) \ra=\partial_x \la u_y^2\partial_x u_y \ra - 2 \la u_y (\partial_x u_y)^2 \ra$. The first term vanishes because of homogeneity, and the second one is zero as well because of axisymmetry: under a rotation of angle $\pi$ around $z$, $u_y$ changes sign but $(\partial_x u_y)^2$ does not. Using the same tricks we show that the coefficient in $r^4$ vanishes, and so does any even order coefficient. We are left with the $r^3$ term only, whose coefficient reads $\la u_y^2 \partial_{xxx}(u_y) \ra / 6 = \la (\partial_x u_y)^3 \ra / 6$. We can write similar expansions for the two other correlation functions $S_{rr\theta}$ and $S_{r\theta r}$. Denoting the horizontal velocity as $\textbf{u}_\perp$ ($x$ and $y$ components of the velocity only) and the vertical vorticity as $\omega$, the result is
\begin{eqnarray}
S_{r r  \theta}(r) & = & \frac{r}{2} \la\textbf{u}_\perp^2 \omega \ra+\frac{r^3}{3} \left( \la (\partial_x u_x)^2  \partial_x u_y\ra + \la u_x \partial_{xx}(u_x) \partial_x u_y \ra  \right) +\mathcal{O}(r^5) \, , \\
S_{r \theta r}(r) & = & -\frac{r}{4} \la\textbf{u}_\perp^2 \omega \ra +\frac{r^3}{6} \left( \frac{1}{2}\la (\partial_x u_x)^2  \partial_x u_y\ra - \la u_x \partial_{xx}(u_x) \partial_x u_y \ra  \right) +\mathcal{O}(r^5) \, , \\
S_{\theta \theta \theta}(r) & = & \frac{r^3}{6}\la (\partial_x u_y)^3 \ra + \mathcal{O}(r^5) \, .
\end{eqnarray}
The correlations $S_{r r  \theta}$ and $S_{ r  \theta r}$ start off linearly in $r$, whereas $S_{\theta \theta \theta}$ increases like $r^3$ close to $r=0$. If we further assume that the flow is two-dimensional, these expressions can be written in terms of the vertical vorticity asymmetry $\la \omega^3 \ra = 2 \la (\partial_x u_y)^3  \ra -6 \la (\partial_x u_y)^2 \partial_y u_x \ra$. Indeed, using the incompressibility constraint for a $z$-independent flow, we obtain $ \la u_x \partial_{xx}(u_x) \partial_x u_y \ra = - \la u_x \partial_{yx}(u_y) \partial_x u_y \ra = - \la u_x \partial_{y}[(\partial_x u_y)^2] \ra /2= \la(\partial_x u_y)^2 \partial_y u_x  \ra /2 $. Upon performing the linear combination $S_{r r  \theta} -4 S_{r \theta r} - S_{\theta \theta \theta}$, we obtain
\begin{equation}
S_{r r  \theta} -4 S_{r \theta r} - S_{\theta \theta \theta}=\frac{3}{2} r \la\textbf{u}_\perp^2 \omega \ra -\frac{r^3}{12} \la \omega^3 \ra + \mathcal{O}(r^5) \, . \label{dev}
\end{equation}
This small-separation expansion highlights the relationship between the antisymmetric velocity correlation functions and the asymmetry of the vorticity distribution, provided the flow is two-dimensional. The typical scale at which the correlations reach an extremal value can be estimated by balancing the two terms on the right-hand-side of (\ref{dev}), which gives
\begin{equation}
r \simeq \sqrt{\frac{\la\textbf{u}_\perp^2 \omega \ra}{ \la \omega^3 \ra }} \, .
\end{equation}


\begin{thebibliography}{99}

\bibitem{CambonBook}
P. Sagaut and C. Cambon, ``Homogeneous turbulence,''
Cambridge University press (2008).

\bibitem{DavidsonBook}
P.A. Davidson, ``Turbulence in rotating, stratified and electrically conducting fluids,''
Cambridge University press (2013).

\bibitem{Cambon1989}
C. Cambon and L. Jacquin, ``Spectral approach to
non-isotropic turbulence subjected to rotation,'' J. Fluid
Mech. {\bf 202}, 295 (1989).  

\bibitem{Staplehurst2008}
P.J. Staplehurst, P.A. Davidson, and S.B. Dalziel, S.B.
``Structure formation in homogeneous freely decaying rotating
turbulence,'' {\it J. Fluid Mech.} \textbf{598}, 81 (2008).  

\bibitem{Davidson2010}
P.A. Davidson, ``On the decay of Saffman turbulence subject
to rotation, stratification or an imposed
magnetic field,'' {\it J. Fluid Mech.} {\bf 663}, 268 (2010). 

\bibitem{Lamriben2011}
C. Lamriben, P.-P. Cortet and F. Moisy, ``Direct measurements of anisotropic energy transfers in a rotating turbulence experiment,'' Phys. Rev. Lett. {\bf 107}, 024503  (2011).

\bibitem{Sen2012}
A. Sen, P.D. Mininni, D. Rosenberg, and A. Pouquet, ``Anisotropy and nonuniversality in scaling laws of the large-scale energy spectrum
in rotating turbulence,'' Phys. Rev. E {\bf 86}, 036319 (2012).

\bibitem{Cortet2013}
P.-P. Cortet and F. Moisy, ``Scale-dependent anisotropy in decaying rotating turbulence,'' subm. to J. Fluid Mech.  (2013).

\bibitem{Hopfinger1982}
E. J. Hopfinger, F. K. Browand, and Y. Gagne, ``Turbulence and waves
in a rotating tank,'' J. Fluid Mech. {\bf 125}, 505 (1982).


\bibitem{Bartello1994}
P. Bartello, O. M\'etais, and M. Lesieur, ``Coherent structures in
rotating
three-dimensional turbulence,'' J. Fluid Mech {\bf 273}, 1 (1994). 

\bibitem{Godeferd1999}
F.S. Godeferd and L. Lollini, ``Direct numerical simulations of
turbulence with confinement and rotation,'' J. Fluid Mech. {\bf 393}, 257 (1999). 

\bibitem{Morize2005} C. Morize, F. Moisy and M. Rabaud, ``Decaying grid-generated turbulence in a rotating tank,'' Phys. Fluids \textbf{17} (9), 095105 (2005).


\bibitem{Bourouiba2007}
L. Bourouiba and P. Bartello, ``The intermediate Rossby number range and two-dimensional three-dimensional transfers in rotating decaying homogeneous turbulence,'' J.
Fluid Mech. {\bf 587}, 139 (2007).  

\bibitem{Sreeni2008}
B. Sreenivasan and P.A. Davidson ``On the formation of cyclones and anticyclones in a rotating fluid,'' {\it Phys. Fluids} {\bf 20}, 085104 (2008).

\bibitem{Bokhoven2008}
L.J.A van Bokhoven, C. Cambon, L. Liechtenstein, F.S. Godeferd, H.J.H.
Clercx, ``Refined vorticity statistics of decaying rotating three-dimensional turbulence,'' J. of Turbulence {\bf 9} (6), 1 (2008). 


\bibitem{Cambon}
C. Cambon and J.F. Scott, ``Linear and nonlinear models of anisotropic turbulence,'' 
Annu. Rev. Fluid Mech. {\bf 31}, 1-53 (1999). 

\bibitem{Moisy2011} F. Moisy, C. Morize, M. Rabaud and J. Sommeria, ``Decay laws, anisotropy and cyclone-anticyclone asymmetry in decaying rotating turbulence,'' J. Fluid Mech. \textbf{664}, 5 (2011).

\bibitem{Frisch}
{U. Frisch},
``Turbulence: the legacy of A.N. Kolmogorov,''
\newblock Cambridge University press (1995).

\bibitem{Gallet}
{B. Gallet and W.R. Young},
``A two-dimensional vortex condensate at high Reynolds number,''
\newblock J. Fluid Mech. \textbf{715}, 359-388 (2013).


\bibitem{Alexakis}
{A. Alexakis and C.R. Doering},
``Energy and enstrophy dissipation in steady state 2d turbulence,''
\newblock Phys. Lett. A, \textbf{359}, 652-657  (2006).

\bibitem{LindborgCho}
{E. Lindborg and J Y. N. Cho},
``Horizontal velocity structure functions in the upper troposphere and lower stratosphere. 2. Theoretical considerations,''
\newblock J. Geophys. Res. \textbf{106}, D10 (2001).


\bibitem{Kloosterziel}
{R.C. Kloosterziel, G.J.F. van Heijst},
``An experimental study of unstable barotropic vortices in a rotating fluid,''
\newblock J. Fluid Mech. \textbf{223}, 1-2, (1991).


\bibitem{Gence}
{J.-N. Gence and C. Frick},
``Birth of the triple correlations of vorticity in an homogeneous turbulence submitted to a solid-body rotation,''
\newblock C. R. Acad. Sci. Paris, \textbf{329}, s\'erie II b, 351-356 (2001).




\bibitem{Billant}
{P. Billant and J.-M. Chomaz},
``Experimental evidence for a new instability of a vertical columnar vortex pair in a strongly stratified fluid,''
\newblock J. Fluid Mech. \textbf{418}, 167-188 (2000).

\bibitem{Augierthese}
{P. Augier},
``Turbulence stratifi\'ee, \'etude des m\'ecanismes de la cascade,''
\newblock Ph.D. thesis, Ecole Polytechnique (2011).

\bibitem{Augier}
{P. Augier, P. Billant, M. E. Negretti, J.-M. Chomaz},
``Experimental study of stratified turbulence forced with columnar dipoles,''
\newblock submitted to Phys. Fluids.


\bibitem{Canuto}
{V.M. Canuto and M.S. Dubovikov},
``Physical regimes and dimensional structures of rotating turbulence,''
\newblock Phys. Rev. Lett. \textbf{78}, 4 (1997).


\bibitem{Mininni}
{P.D. Mininni, D. Rosenberg and A. Pouquet},
``Isotropization at small scales of rotating helically-driven turbulence,''
\newblock J. Fluid Mech. \textbf{699}, 263-279 (2012).




\bibitem{Chavanis}
{P.-H. Chavanis and C. Sire},
``The spatial correlations in the velocities arising from a random distribution of point vortices,''
\newblock Phys. Fluids, \textbf{13}, 1904 (2001).

\bibitem{Townsend}
{A. A. Townsend},
``On the fine-scale structure of Turbulence,''
\newblock Proc. R. Soc. London, Ser. A, \textbf{238}, 534 (1951).

\bibitem{Lundgren}
{T. S. Lundgren},
``Strained spiral vortex model for turbulent fine structure,''
\newblock Phys. Fluids, \textbf{25}, 2193 (1982).


\bibitem{Saffman}
{P. G. Saffman and D.I. Pullin},
``Calculation of velocity structure functions for vortex models of isotropic turbulence,''
\newblock Phys. Fluids, \textbf{8}, 11 (1996).


\bibitem{Hatakeyama}
{N. Hatakeyama and T. Kambe},
``Statistical laws of random strained vortices in turbulence,''
\newblock Phys. rev. Lett., \textbf{79}, 7 (1997).

\bibitem{He}
{G. He, G. D. Doolen, S. Chen},
``Calculations of longitudinal and transverse velocity structure functions using a vortex model of isotropic turbulence,''
\newblock Phys. Fluids, \textbf{11}, 12 (1999).


\bibitem{Lindborg}
{E. Lindborg},
``Kinematics of homogeneous axisymmetric turbulence,''
\newblock J. Fluid Mech., \textbf{302}, 179-201 (1995).



\bibitem{Oughton}
{S. Oughton, H.-H. R\"adler, W.H. Matthaeus},
``General second-rank correlation tensors for homogeneous magnetohydrodynamic turbulence,''
\newblock Phys. Rev. E, \textbf{56}, 3 (1997).


\end{thebibliography}
\end{document}